\documentclass[prx, twocolumn, superscriptaddress]{revtex4-2}
\usepackage{bm, amsmath, amsfonts, amssymb, mathtools, braket}
\usepackage{times}
\usepackage{multirow}
\usepackage{graphicx}
\usepackage{float, color}

\usepackage[
pagebackref=false,
colorlinks=true,
linkcolor=blue,
urlcolor=blue,
filecolor=black,
citecolor=red,
pdfstartview=FitV,
pdftitle={},
pdfauthor={},
pdfsubject={},
pdfkeywords={},
pdfpagemode=None,
bookmarksopen=true
]{hyperref}

\begin{document}

\newcommand{\ii}{\text{i}}
\newcommand{\vac}{\rm vac}

\newcommand{\magenta}[1]{{\textcolor{magenta}{#1}}}

\title{Entanglement Phase Transition Induced by the Non-Hermitian Skin Effect}

\author{Kohei Kawabata}
\email{kohei.kawabata@princeton.edu}
\affiliation{Department of Physics, Princeton University, Princeton, New Jersey 08544, USA}

\author{Tokiro Numasawa}
\affiliation{Institute for Solid State Physics, University of Tokyo, Kashiwa 277-8581, Japan}

\author{Shinsei Ryu}
\affiliation{Department of Physics, Princeton University, Princeton, New Jersey 08544, USA}

\date{\today}

\begin{abstract}
Recent years have seen remarkable development in open quantum systems effectively described by non-Hermitian Hamiltonians. 
A unique feature of non-Hermitian topological systems is the skin effect, anomalous localization of an extensive number of eigenstates driven by nonreciprocal dissipation.
Despite its significance for non-Hermitian topological phases, the relevance of the skin effect to quantum entanglement and critical phenomena has remained unclear.
Here, we find that the skin effect induces a nonequilibrium quantum phase transition in the entanglement dynamics.
We show that the skin effect gives rise to a macroscopic flow of particles and suppresses the entanglement propagation and thermalization, leading to the area law of the entanglement entropy in the nonequilibrium steady state.
Moreover, we reveal an entanglement phase transition induced by the competition between the unitary dynamics and the skin effect even without disorder or interactions.
This entanglement phase transition accompanies nonequilibrium quantum criticality characterized by a nonunitary conformal field theory whose effective central charge is extremely sensitive to the boundary conditions.
We also demonstrate that it originates from an exceptional point of the non-Hermitian Hamiltonian and the concomitant scale invariance of the skin modes localized according to the power law.
Furthermore, we show that the skin effect leads to the purification and the reduction of von Neumann entropy even in Markovian open quantum systems described by the Lindblad master equation.
Our work opens a way to control the entanglement growth and establishes a fundamental understanding of phase transitions and critical phenomena in open quantum systems far from thermal equilibrium. 
\end{abstract}

\maketitle


\section{Introduction}

Nonequilibrium quantum dynamics provides a profound understanding about 
quantum many-body systems.
Closed quantum systems driven out of equilibrium eventually reach thermal equilibrium, which validates the foundations of quantum statistical mechanics~\cite{Polkovnikov-review, Eisert-review, Huse-review, Rigol-review}.
Thanks to the recent advances in quantum simulations and technologies, such thermalization dynamics was experimentally observed in ultracold atoms~\cite{Trotzky-12, Gring-12, Kaufman-16} 
and trapped ions~\cite{Smith-16}.
Thermalization arises from the propagation of quantum correlations and entanglement throughout the whole system and the consequent entanglement entropy proportional to the volume of the subsystem~\cite{Calabrese-Cardy-05, *Calabrese-Cardy-06, Kim-13, Nahum-17}.
Beyond closed quantum systems, the nonequilibrium dynamics of open quantum systems has recently been studied extensively.
Researchers have found entanglement phase transitions induced by quantum measurements~\cite{Chan-19, Skinner-19, Li-18, *Li-19, Choi-20, *Bao-20, Gullans-20, Cao-19, Jian-20, Lavasani-21, Sang-21, Ippoliti-21, Fuji-20, Alberton-21, Ippoliti-21-spacetimeL, *Ippoliti-21-spacetimeX, Lu-21}.
There, sufficiently strong quantum measurements prevent thermalization and drive the system into a steady state far from equilibrium for which the entanglement entropy is only proportional to the boundary of the subsystem
(i.e., the area law~\cite{Eisert-area-review}).
Such measurement-induced phase transitions also accompany nonequilibrium critical phenomena unique to open quantum systems.

As another platform of open systems, the physics effectively described by non-Hermitian Hamiltonians has recently attracted growing interest~\cite{Konotop-review, Christodoulides-review}.
In the classical regime, non-Hermiticity is implemented by controlling gain and loss, and leads to unique phenomena and functionalities without Hermitian counterparts, 
such as power oscillations~\cite{Makris-08, Ruter-10, Regensburger-12}, 
unidirectional invisibility~\cite{Mostafazadeh-09, Lin-11, Feng-13, Peng-14NP}, 
high-performance lasers~\cite{Longhi-10, Chong-11, Peng-14S, Feng-14, Hodaei-14},
and enhanced sensitivity~\cite{Wiersig-14, Hodaei-17, Chen-17}. 
In the quantum regime, effective non-Hermitian Hamiltonians are justified as conditional dynamics subject to continuous monitoring and postselection of the null measurement outcome~\cite{Dalibard-92, *Molmer-93, Dum-92, Carmichael-textbook, Plenio-review, Daley-review}, as well as the Feshbach projection formalism~\cite{Gamow-28, Feshbach-54, *Feshbach-58, *Feshbach-62, Rotter-review, Moiseyev-textbook}. 
Non-Hermitian systems have been realized in several open quantum systems, including atoms~\cite{Peng-16, Li-16, Ren-22}, 
photons~\cite{Xiao-17, Kawabata-17, *Xiao-19, Dora-19, *Xiao-PRXQuantum21, Ozturk-21}, 
exciton-polaritons~\cite{Gao-15}, 
electronic spins~\cite{Wu-19, Zhang-Duan-21},
and superconducting qubits~\cite{Naghiloo-19}. 
On the theoretical side, researchers have studied open quantum dynamics of non-Hermitian systems~\cite{Bender-07, Brody-12, Ashida-18, Dora-20, *Bacsi-21, Chen-20, Biella-21, *Turkeshi-22, Gopalakrishnan-21, Jian-21, Orito-22}.
Notably, non-Hermitian systems at critical points support anomalous singularities called exceptional points~\cite{Kato-textbook, Berry-04, Heiss-12}, at which the non-Hermitian Hamiltonians are no longer diagonalizable.
Phase transitions and critical phenomena due to exceptional points date back to the Yang-Lee edge singularity~\cite{Yang-Lee-52I, *Lee-Yang-52II, Fisher-78, Cardy-85, Itzykson-Drouffe-textbook}.
Exceptional points are also the key to the real-complex spectral transition protected by parity-time symmetry~\cite{Bender-02, Bender-review} and induce new universality classes of phase transitions in non-Hermitian quantum systems~\cite{Fukui-98PRB, Nakamura-06, Oka-10, Lee-14X, *Lee-14L, Ashida-17, Couvreur-17, Nakagawa-18, Herviou-19, *Herviou-19-ES, Yamamoto-19, Chang-20, *Tu-21, CHLee-22}.

Another unique feature of non-Hermitian systems is the skin effect~\cite{Lee-16, YW-18-SSH, *YSW-18-Chern, Kunst-18}.
This is anomalous localization of an extensive number of eigenstates driven by reciprocity-breaking non-Hermiticity, which has no analogs in Hermitian systems.
The skin effect plays a central role in the topological phases of non-Hermitian systems~\cite{Rudner-09, Sato-11, *Esaki-11, Hu-11, Schomerus-13, Longhi-15, Leykam-17, Xu-17, Shen-18, *Kozii-17, Gong-18, *Kawabata-19, KSUS-19, ZL-19, Zirnstein-19, Borgnia-19, KBS-19, JYLee-19, KSR-21, Bergholtz-review}.
Since the skin effect leads to extreme sensitivity of the bulk to the boundary conditions, it changes the nature of the bulk-boundary correspondence~\cite{Lee-16, Xiong-18, MartinezAlvarez-18, YW-18-SSH, *YSW-18-Chern, Kunst-18, McDonald-18, Lee-Thomale-19, Liu-19, Lee-Li-Gong-19, Yokomizo-19, Okuma-21}.
Moreover, the skin effect originates from the topological invariants intrinsic to non-Hermitian systems~\cite{Zhang-20, OKSS-20, *KOS-20, KSR-21}.
The skin effect has recently been observed in classical experiments of mechanical metamaterials~\cite{Brandenbourger-19-skin-exp, *Ghatak-19-skin-exp},
electrical circuits~\cite{Helbig-19-skin-exp, *Hofmann-19-skin-exp, Zhang-21}, 
photonic lattices~\cite{Weidemann-20-skin-exp}, 
and active particles~\cite{Palacios-21},
as well as quantum experiments of single photons~\cite{Xiao-19-skin-exp} and ultracold atoms~\cite{Gou-20, *Liang-22}.
In these experiments, reciprocity-breaking dissipation is introduced by the asymmetry of the hopping amplitudes.
It is also relevant to Liouvillians for a quantum master equation~\cite{Song-19, Haga-21, Liu-20PRR, Mori-20, Yang-22}.
The skin effect may open up a way to actively control the phases of matter.

Despite the significance of the skin effect for non-Hermitian topological phases, its impact on the genuine quantum nature has remained unclear.
While several recent works studied the entanglement dynamics in non-Hermitian quantum systems~\cite{Ashida-18, Dora-20, *Bacsi-21, Chen-20, Biella-21, Gopalakrishnan-21, Jian-21, Orito-22}, they focused only on non-Hermitian systems that are subject to reciprocal dissipation and free from the skin effect.
On the basis of the important role of the skin effect in non-Hermitian physics, it may crucially change the entanglement dynamics in open quantum systems.
Furthermore, the relevance of the skin effect on quantum phase transitions has also been unclear.
The previous works focused on the Yang-Lee edge singularity~\cite{Yang-Lee-52I, *Lee-Yang-52II, Fisher-78, Cardy-85, Itzykson-Drouffe-textbook} and its variants~\cite{Lee-14X, *Lee-14L, Ashida-17, Couvreur-17, Kawabata-17, Nakagawa-18, Dora-20, Chang-20, CHLee-22}, which do not accompany the skin effect.
Although the skin effect may lead to new universality classes of phase transitions and critical phenomena far from thermal equilibrium, no research has hitherto addressed this problem.

In this work, we study the impact of the skin effect on the entanglement dynamics and nonequilibrium phase transitions in open quantum systems.
First, we show that the skin effect gives rise to a macroscopic flow of particles and suppresses the entanglement propagation, leading to a nonequilibrium steady state characterized by the area law of entanglement entropy.
This is contrasted with the thermal equilibrium states, which exhibit the volume law of entanglement entropy.
Second, we reveal a new type of entanglement phase transition induced by the skin effect.
It arises from the competition between coherent coupling and nonreciprocal dissipation; 
the nonequilibrium steady state exhibits the volume law for small dissipation but the area law for large dissipation, between which the entanglement entropy grows subextensively (i.e., logarithmically with respect to the subsystem size).
Anomalously, this nonequilibrium quantum criticality is characterized by a nonunitary conformal field theory whose effective central charge is extremely sensitive to the boundary conditions.
We also demonstrate that it originates from an exceptional point in the non-Hermitian Hamiltonian and the concomitant scale 
invariance of the skin modes localized according to the power law.
In addition to the conditional dynamics effectively described by non-Hermitian Hamiltonians, we show that the skin effect leads to the purification and the reduction of von Neumann entropy even in Markovian open quantum systems described by the Lindblad master equation.

From these results, we show that the skin effect is a new mechanism that triggers entanglement phase transitions and nonequilibrium critical phenomena in open quantum systems.
The measurement-induced phase transitions typically rely on spatial or temporal randomness~\cite{Chan-19, Skinner-19, Li-18, *Li-19, Choi-20, *Bao-20, Gullans-20, Cao-19, Jian-20, Lavasani-21, Sang-21, Ippoliti-21, Fuji-20, Alberton-21, Ippoliti-21-spacetimeL, *Ippoliti-21-spacetimeX, Lu-21} while they can occur in some models with no randomness except in measurement outcomes~\cite{Li-19}.
The entanglement phase transition in this work relies not on any randomness but on the skin effect.
While the Yang-Lee edge singularity~\cite{Yang-Lee-52I, *Lee-Yang-52II, Fisher-78, Cardy-85, Itzykson-Drouffe-textbook} originates from an exceptional point, it does not accompany the skin effect.
Furthermore, the boundary-sensitive effective central charge, which implies a new universality class, has never been reported in conformal field theory.
Since the skin effect is a universal phenomenon arising solely from non-Hermitian topology, our entanglement phase transition can generally appear in a wide variety of open quantum systems.
We hope that these results will deepen our understanding of quantum phases far from thermal equilibrium.

The rest of this work is organized as follows.
In Sec.~\ref{sec: entanglement-skin}, we describe general behavior of the entanglement dynamics in closed and open quantum systems.
In Sec.~\ref{sec: HN}, we show the entanglement suppression induced by the skin effect for a non-Hermitian spinless-fermionic model.
In Sec.~\ref{sec: sHN}, we demonstrate the entanglement phase transition and discuss its nonequilibrium quantum criticality for a non-Hermitian spinful-fermionic model.
In Sec.~\ref{sec: Liouvillian}, we show that the skin effect leads to the purification and reduction of von Neumann entropy in a Liouvillian of the Lindblad master equation.
In Sec.~\ref{sec: conclusion}, we conclude this work with several outlooks.
In Appendix~\ref{asec: NH}, we describe the implementation of effective non-Hermitian Hamiltonians in the quantum trajectory approach.
In Appendix~\ref{asec: numerics}, we describe the numerical method to effectively simulate the dynamics of non-Hermitian free fermions.
In Appendix~\ref{asec: initial conditions}, we provide additional numerical results for different initial conditions.
In Appendix~\ref{asec: Liouvillian}, we describe details of the Liouvillian skin effect.

\section{Entanglement dynamics and non-Hermitian skin effect}
    \label{sec: entanglement-skin}

\begin{figure}[t]
\centering
\includegraphics[width=0.85\linewidth]{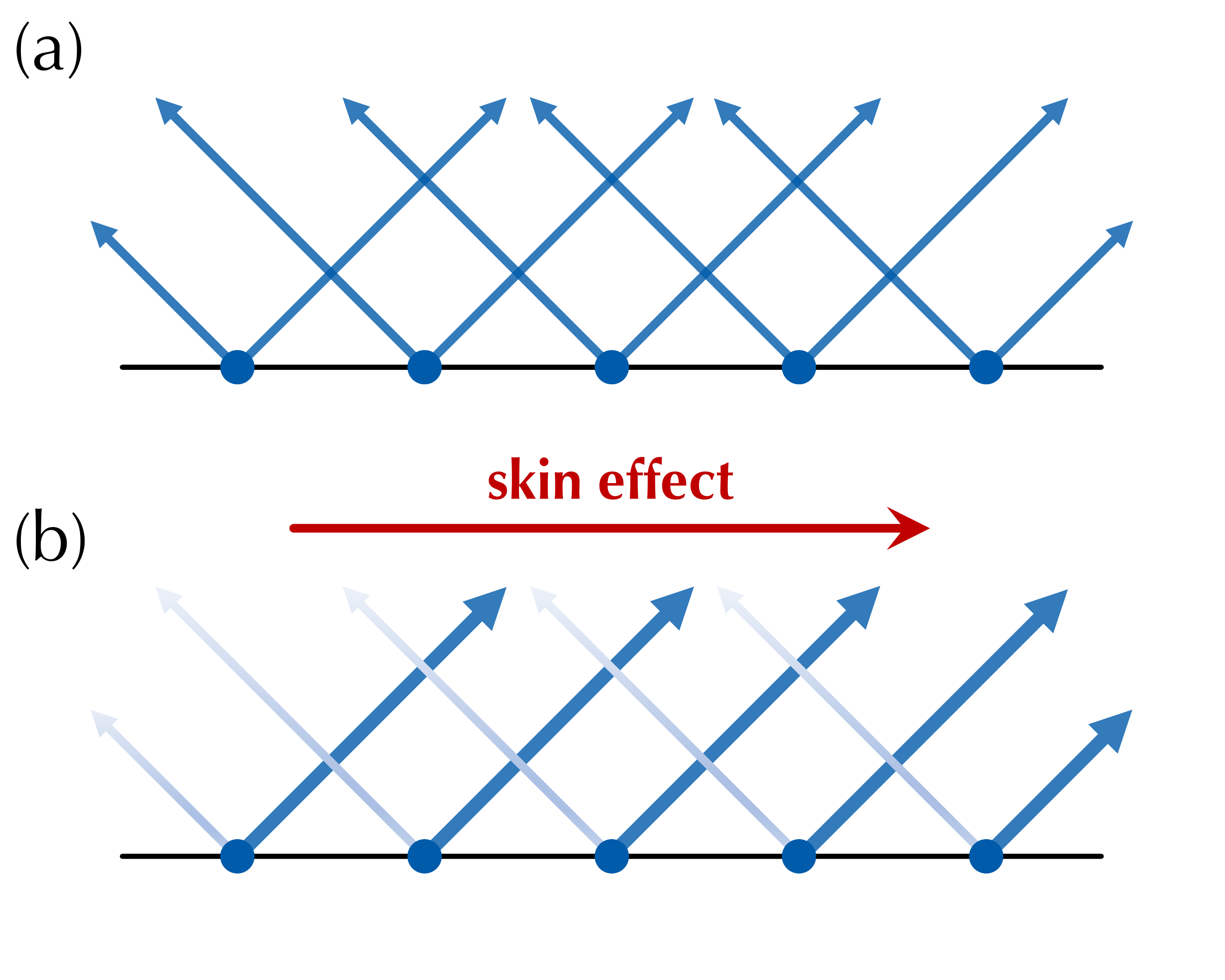} 
\caption{Quasiparticle propagation in closed and open quantum systems. (a)~Closed quantum systems. Quasiparticles propagate in both directions and diffuse throughout the system, leading to the volume law of entanglement entropy. (b)~Open quantum systems subject to the skin effect. Nonreciprocal dissipation makes quasiparticles move toward only one direction, suppressing the entanglement propagation and leading to the area law of entanglement entropy.}
	\label{fig: schematic}
\end{figure}

Before the detailed calculations, we discuss the general behavior of nonequilibrium dynamics in closed and open quantum systems.
For simplicity, we assume
the quasiparticle picture, which is applicable to integrable systems discussed in this work.
Under the time evolution of closed quantum systems, the quasiparticles coherently move in all the directions and diffuse throughout the entire system [Fig.~\ref{fig: schematic}\,(a)].
Such a bidirectional propagation of quasiparticles arises from the conservation of the particle number and energy.
Consequently, quantum correlations develop throughout the system, leading to extensive entanglement for the steady state.
This means the entanglement entropy proportional to the subsystem
size, i.e., volume law ($S \propto l^d$ with the subsystem length $l$ and spatial dimensions $d$)~\cite{Calabrese-Cardy-05, *Calabrese-Cardy-06}.
The volume law of the entanglement entropy lies at the heart of thermalization and validates quantum statistical mechanics~\cite{Polkovnikov-review, Eisert-review, Huse-review, Rigol-review}.

In open quantum systems, the particle number or energy is not necessarily conserved because of the coupling to the external environment.
As a direct result of the violation of the conservation laws, quasiparticles can be amplified or attenuated.
As long as such an external coupling is reciprocal, quantum correlations propagate uniformly throughout the system in a manner similar to closed quantum systems.
However, when the external coupling is nonreciprocal, quasiparticles can be amplified toward one direction and attenuated toward the other direction [Fig.~\ref{fig: schematic}\,(b)].
In such a case, the quasiparticles move only in one direction and accumulate at a boundary for a sufficiently long time, i.e., non-Hermitian skin effect~\cite{Lee-16, YW-18-SSH, Kunst-18}.
Since the quasiparticles are present only at a boundary, the quantum correlations extend not over the entire system but only at the boundary.
The entanglement is greatly suppressed and
carried  
only by the skin modes at the boundary, leading to the area law of the entanglement entropy (i.e., $S \propto l^{d-1}$).
This is a unique consequence of nonreciprocal dissipation for quantum entanglement dynamics.
We confirm such a suppression of entanglement for a non-Hermitian spinless-fermionic model (i.e., Hatano-Nelson model~\cite{Hatano-Nelson-96, *Hatano-Nelson-97}) in Sec.~\ref{sec: HN}.

Notably, an extensive number of localized modes are needed for the entanglement suppression.
A possible known mechanism that gives rise to it is disorder. 
In the presence of sufficiently strong disorder, the system is subject to the Anderson~\cite{Lee-review, Evers-review} or many-body~\cite{Huse-review} localization, in which thermalization is prohibited.
We emphasize that the skin effect is a different mechanism that suppresses the entanglement growth.
In fact, the skin effect does not rely on disorder, and occurs only in open quantum systems.
The skin effect originates solely from non-Hermitian topology~\cite{Zhang-20, OKSS-20, *KOS-20, KSR-21} and hence appears in a wide variety of open quantum systems.

Even if the skin effect suppresses the quasiparticle diffusion and the entanglement propagation, it is unclear whether the skin effect can compete with the unitary dynamics and give rise to a continuous phase transition.
In fact, in the Hatano-Nelson model and many other non-Hermitian models, even infinitesimal non-Hermiticity causes the skin effect and results in no continuous phase transition.
Nevertheless, we show that the skin effect indeed induces new nonequilibrium phase transitions and critical phenomena intrinsic to open quantum systems.
There, an entanglement phase transition arises from the competition between the coherent coupling and the nonreciprocal dissipation:
the system reaches a thermal equilibrium state exhibiting the volume law for small dissipation while it reaches a nonequilibrium steady state exhibiting only the area law for large dissipation, between which the entanglement entropy grows subextensively (i.e., $S \propto \log l$) with an unconventional nonequilibrium quantum criticality described by a nonunitary conformal field theory.
We demonstrate such an entanglement phase transition induced by the skin effect by explicitly constructing and investigating an illustrative example of non-Hermitian spinful-fermionic models (i.e., symplectic Hatano-Nelson model~\cite{OKSS-20, *KOS-20, KR-21}) in Sec.~\ref{sec: sHN}.

As well as the conditional dynamics effectively described by non-Hermitian Hamiltonians, the skin effect has a considerable impact also on the open quantum dynamics described by a master equation.
While a Markovian open quantum system typically exhibits the thermal equilibrium state with infinite temperature as the steady state, the skin effect dramatically changes the properties of the steady state toward far from equilibrium.
We show the purification and suppression of von Neumann entropy for Markovian open quantum systems described by the Lindblad master equation in Sec.~\ref{sec: Liouvillian}.

Entanglement phase transitions can also occur as a consequence of the competition between the unitary dynamics and quantum measurements~\cite{Chan-19, Skinner-19, Li-18, *Li-19, Choi-20, *Bao-20, Gullans-20, Cao-19, Jian-20, Lavasani-21, Sang-21, Ippoliti-21, Fuji-20, Alberton-21, Ippoliti-21-spacetimeL, *Ippoliti-21-spacetimeX, Lu-21}.
However, the entanglement phase transition in this work exhibits properties distinct from the measurement-induced phase transitions.
First, the boundary-sensitive critical behaviors have never been found in the previous works on the measurement-induced phase transitions.
Additionally, the measurement-induced phase transitions typically rely on spatial or temporal randomness and many-body interactions aside from some exceptions~\cite{Li-19, Lu-21}.
By contrast, the skin effect induces the entanglement phase transition even without randomness and interactions, which enables a deep understanding of the phase transition and critical behavior in open quantum systems.
Furthermore, the measurement-induced phase transitions manifest themselves only in a conditional quantum trajectory postselected by measurements and disappear in the open quantum dynamics averaged over multiple quantum trajectories.
On the other hand, the skin effect occurs and yields purification even in the averaged open quantum dynamics described by the Markovian master equation.

\section{Entanglement suppression induced by the non-Hermitian skin effect}
    \label{sec: HN}

We study the nonequilibrium quantum dynamics induced by the non-Hermitian skin effect.
To this end, we investigate the Hatano-Nelson model~\cite{Hatano-Nelson-96, *Hatano-Nelson-97} as a prototypical example that exhibits the skin effect:
\begin{equation}
    \hat{H} = - \frac{1}{2} \sum_{l} \left[ \left( J+\gamma \right) \hat{c}_{l+1}^{\dag} \hat{c}_{l} 
    + \left( J-\gamma \right) \hat{c}_{l}^{\dag} \hat{c}_{l+1} \right],
        \label{eq: HN}
\end{equation}
where $\hat{c}_{l}$ ($\hat{c}_{l}^{\dag}$) annihilates (creates) a spinless fermion at site $l$, $J > 0$ denotes the Hermitian hopping amplitude, and $\gamma \in \mathbb{R}$ denotes the asymmetric hopping amplitude as a source
of non-Hermiticity. 
Here, we assume $\left| \gamma \right| < J$ for simplicity.
The asymmetric hopping can be implemented in the quantum trajectory approach (see Appendix~\ref{asec: NH} for details)~\cite{Dalibard-92, *Molmer-93, Dum-92, Carmichael-textbook, Plenio-review, Daley-review} and has been realized in the recent experiments of single photons~\cite{Xiao-19-skin-exp} and ultracold atoms~\cite{Gou-20, *Liang-22}.

Under the periodic boundary conditions, the Bloch Hamiltonian for the Hatano-Nelson model reads
\begin{equation}
    H \left( k \right) = -J \cos k + \ii \gamma \sin k.
\end{equation}
Thus, the complex-valued spectrum of $H \left( k \right)$ winds around the origin in the complex-energy plane when the momentum $k$ goes around the Brillouin zone $\left[ 0, 2\pi \right)$.
From this complex-spectral winding, we introduce a topological invariant~\cite{Gong-18, KSUS-19} 
\begin{equation}
    W \coloneqq - \oint_{0}^{2\pi} \frac{dk}{2\pi\ii} \frac{d}{dk} \log \det H \left( k \right).
\end{equation}
Since such complex-spectral winding is ill defined in Hermitian systems, the winding number $W$ is intrinsic to non-Hermitian systems.
As a consequence of the intrinsic non-Hermitian topology, an extensive number of boundary modes 
appear
under the open boundary conditions~\cite{Zhang-20, OKSS-20}, i.e., non-Hermitian skin effect~\cite{Lee-16, YW-18-SSH, *YSW-18-Chern, Kunst-18}.
While we here focus on the Hatano-Nelson model in Eq.~(\ref{eq: HN}) as a prototypical example, the skin effect generally occurs and leads to the entanglement suppression as long as the intrinsic non-Hermitian topology is nontrivial $W \neq 0$.

In the following, we impose the open boundary conditions and prepare the initial state as the
charge density wave state
\begin{equation}
    \ket{\psi_0} = \left( \prod_{l=1}^{L/2} \hat{c}_{2l}^{\dag} \right) \ket{\rm vac},
        \label{eq: HN-CDW}
\end{equation}
where $\ket{\rm vac}$ is the fermionic vacuum state, and the system length $L$ is assumed to be even.
The many-particle wave function evolves by the non-Hermitian Hamiltonian $\hat{H}$ in Eq.~(\ref{eq: HN}) as
\begin{equation}
    \ket{\psi \left( t \right)} = \frac{e^{-\ii \hat{H} t} \ket{\psi_0}}{\| e^{-\ii \hat{H} t} \ket{\psi_0} \|}.
        \label{eq: NH-dynamics}
\end{equation}
Despite non-Hermiticity of the Hamiltonian, the particle number $N = L/2$ is conserved under dynamics.
Thanks to the free (i.e., quadratic) nature of the model, its dynamics can be efficiently calculated (see Appendix~\ref{asec: numerics} for details).
We show that the skin effect leads to a nonequilibrium steady state whose entanglement is suppressed, which is to be contrasted with the thermal equilibrium states in closed quantum systems.
While we here consider Eq.~(\ref{eq: HN-CDW}) as an initial state, the entanglement suppression depends only on the skin effect, and the specific details of the initial state should be irrelevant.

\subsection{Skin effect}
    \label{sec: HN-skin}

\begin{figure}[t]
\centering
\includegraphics[width=\linewidth]{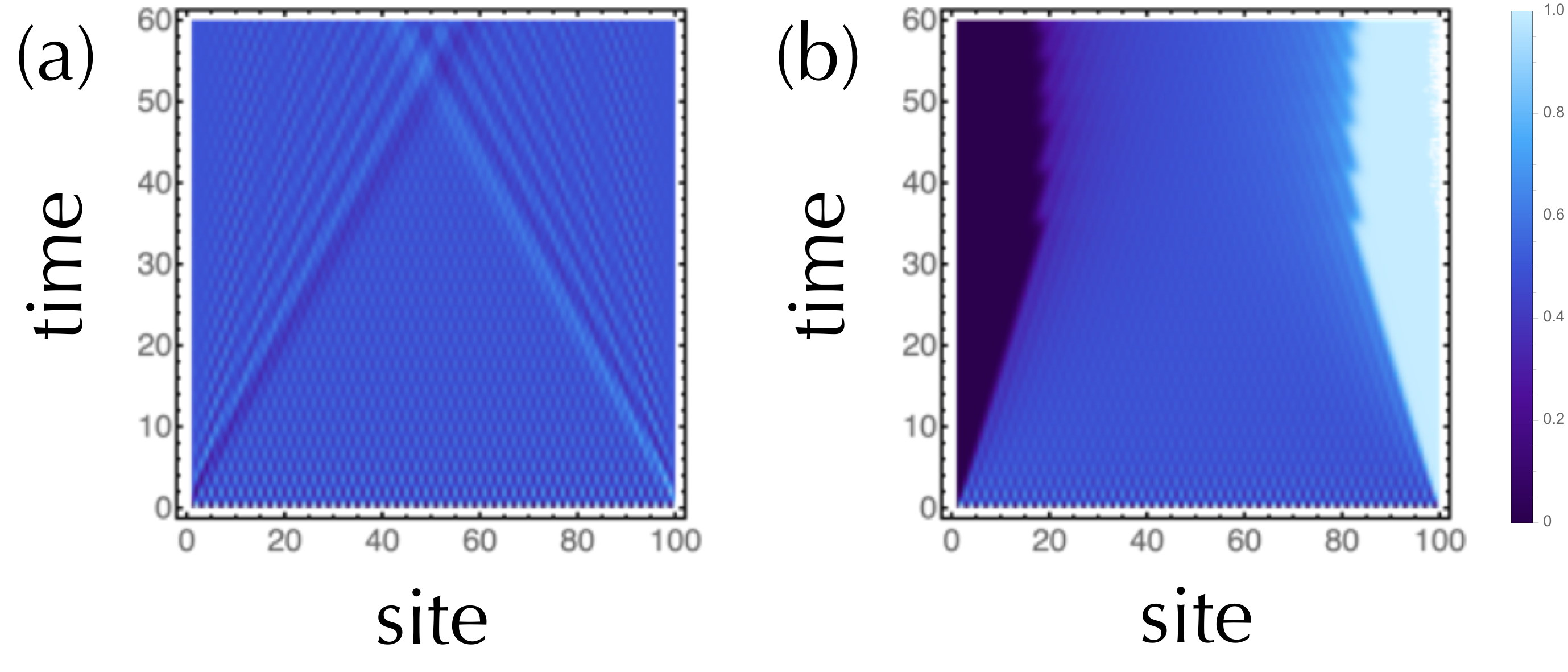} 
\caption{Time evolution of the local particle number $n_{l} \left( t \right) \coloneqq \braket{\psi \left( t \right) | \hat{n}_{l} | \psi \left( t \right)}$ in the Hatano-Nelson model with open boundaries ($L=100$, $J=1.0$) for (a)~$\gamma = 0.0$ and (b)~$\gamma = 0.8$. 
The initial state is prepared as the charge density wave state in Eq.~(\ref{eq: HN-CDW}).
In the presence of non-Hermiticity, particles accumulate at the boundary, which is a clear signature of the non-Hermitian skin effect.}
	\label{fig: HN-number}
\end{figure}

We begin with investigating the time evolution of the local particle number
\begin{equation}
    n_{l} \left( t \right) \coloneqq \braket{\psi \left( t \right) | \hat{n}_{l} | \psi \left( t \right)}.
        \label{eq: HN - local particle number}
\end{equation}
In Hermitian systems, particles are 
distributed uniformly
[Fig.~\ref{fig: HN-number}\,(a)].
In the presence of non-Hermiticity, by contrast, particles accumulate at the right (left) edge of the system for $\gamma > 0$ ($\gamma < 0$) [Fig.~\ref{fig: HN-number}\,(b)].
Such localization of an extensive number of particles is impossible in closed quantum systems and is a clear signature of the non-Hermitian skin effect.

The skin effect can be understood by the imaginary gauge transformation ($\mathrm{GL} \left( 1 \right)$ gauge transformation; $\mathrm{GL} \left( n \right)$ is the general linear group of $n \times n$ invertible matrices)~\cite{Hatano-Nelson-96, *Hatano-Nelson-97, YW-18-SSH, Yokomizo-19}.
Let us introduce the new fermionic operators by
\begin{equation}
    \hat{p}_{l}^{\dag} \coloneqq e^{l\theta} \hat{c}_{l}^{\dag},
    \quad
    \hat{q}_{l} \coloneqq e^{-l\theta} \hat{c}_{l}
        \label{eq: GL1 gauge}
\end{equation}
where $\theta \in \mathbb{C}$ plays a role of the complex-valued gauge.
The Hamiltonian in Eq.~(\ref{eq: HN}) is rewritten as
\begin{equation}
    \hat{H} = - \frac{1}{2} \sum_{l=1}^{L-1} \left[ e^{-\theta }\left( J+\gamma \right) \hat{p}_{l+1}^{\dag} \hat{q}_{l} 
    + e^{\theta} \left( J-\gamma \right) \hat{p}_{l}^{\dag} \hat{q}_{l+1} \right].
\end{equation}
In particular, when we choose $\theta$ so that it will satisfy $e^{-\theta }\left( J+\gamma \right) = e^{\theta} \left( J-\gamma \right)$, i.e., 
\begin{equation}
    \theta = \frac{1}{2} \log \left( \frac{J+\gamma}{J-\gamma} \right),
\end{equation}
the Hamiltonian reduces to
\begin{equation}
    \hat{H} = - \frac{\sqrt{J^2-\gamma^2}}{2} \sum_{l=1}^{L-1} \left( \hat{p}_{l+1}^{\dag} \hat{q}_{l} 
    + \hat{p}_{l}^{\dag} \hat{q}_{l+1} \right).
\end{equation}
Now that the asymmetric hopping formally disappears, the Hamiltonian is diagonalized to
\begin{equation}
    \hat{H} = - \sqrt{J^2 - \gamma^2} \sum_{k} \left( \cos k \right) \hat{p}_{k}^{\dag} \hat{q}_{k}
        \label{eq: HN-OBC-diagonalization}
\end{equation}
by the Fourier transforms
\begin{align}
    \hat{p}_{k} &\coloneqq \sqrt{\frac{2}{L+1}} \sum_{l=1}^{L} \hat{p}_{l} \sin \left( kl \right), \\
    \hat{q}_{k} &\coloneqq \sqrt{\frac{2}{L+1}} \sum_{l=1}^{L} \hat{q}_{l} \sin \left( kl \right),
\end{align}
with momentum $k = n\pi/\left( L+1 \right)$ ($n=1, 2, \cdots, L$).
Thus, the spectrum of $\hat{H}$ is entirely real.
Non-Hermiticity of $\hat{H}$ originates solely from the nonorthogonality of the quasiparticles (i.e., $\hat{p}_{k} \neq \hat{q}_{k}$).
In the presence of the skin effect,
while the spectrum of an infinite non-Hermitian system coincides with the infinite-size limit of the spectrum of the corresponding finite system with periodic boundaries,
it does not coincide with the spectrum of the infinite-size limit of the corresponding finite system with open boundaries~\cite{OKSS-20, Trefethen-Embree-textbook}.
This extreme sensitivity yields unique open quantum phenomena, as shown below.

Because of the $\mathrm{GL} \left( 1 \right)$ transformation in Eq.~(\ref{eq: GL1 gauge}), the quasiparticle $\hat{p}_k$ is exponentially localized at the right (left) edge while the quasiparticle $\hat{q}_{k}$ is exponentially localized at the left (right) edge for $\mathrm{Re}\,\theta > 0$ ($\mathrm{Re}\,\theta < 0$).
All the quasiparticles are localized at the edges, which is the hallmark of the skin effect unique to non-Hermitian systems.
Thus, the Hamiltonian $\hat{H}$ annihilates the quasiparticles around one edge and creates the quasiparticles around the other edge under its time evolution.
Here, $\theta^{-1}$ characterizes the localization length of the quasiparticles.
It should be noted that the above transformation is possible only for the open boundary conditions and is unfeasible so that the periodic boundary conditions can be satisfied.
The quasiparticles form Bloch waves delocalized throughout the system under the periodic boundary conditions, where no length scale appears as a consequence of non-Hermiticity.
The emergent length scale $\theta^{-1}$ is unique to the open boundary conditions.

\begin{figure}[t]
\centering
\includegraphics[width=\linewidth]{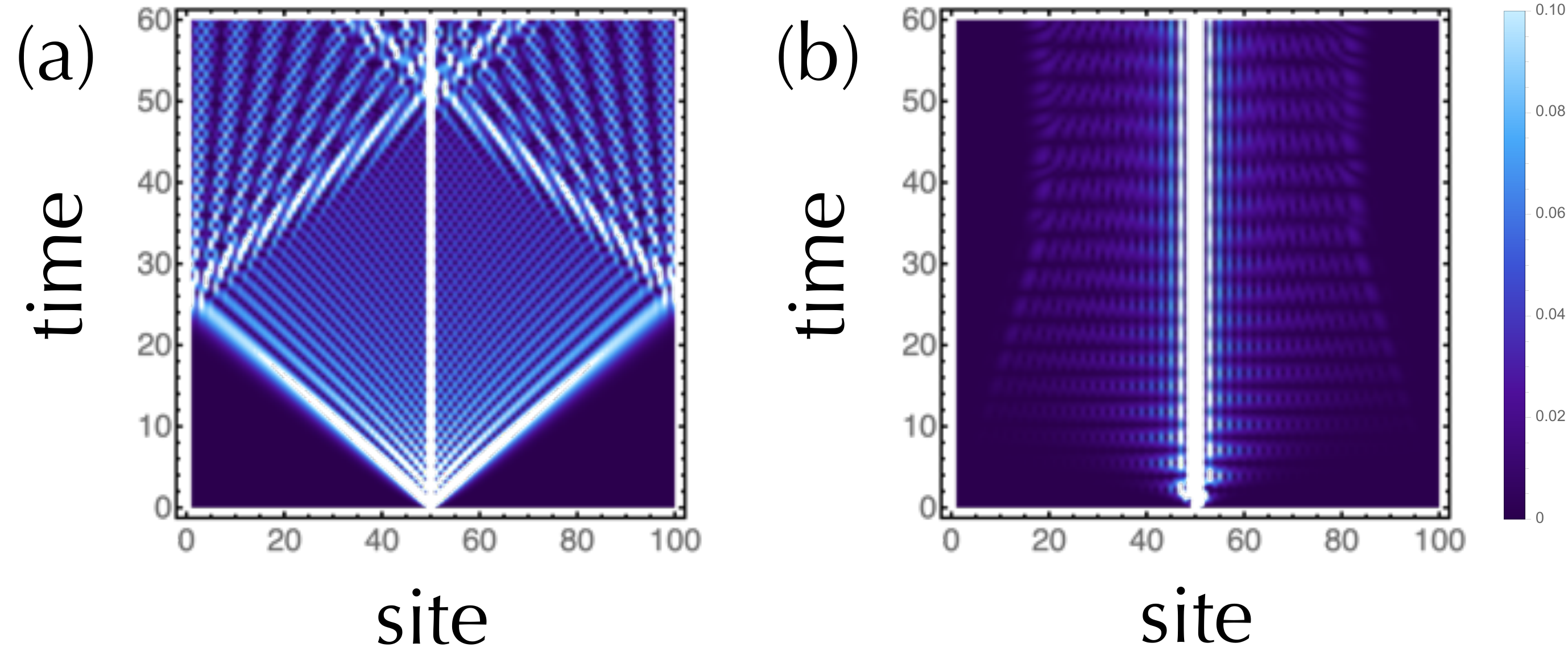} 
\caption{Correlation propagation in the Hatano-Nelson model  with open boundaries ($L=100$, $J=1.0$) for (a)~$\gamma = 0.0$ and (b)~$\gamma = 0.8$. 
The absolute values $\left| C_{l, l_0} \right|$ of the correlation matrix are shown as a function of site $l$ and time $t$ with $l_0 = L/2 = 50$. 
The initial state is prepared as the charge density wave state in Eq.~(\ref{eq: HN-CDW}).
In the presence of non-Hermiticity, the correlation propagation is frozen as a consequence of the non-Hermitian skin effect.}
	\label{fig: HN-correlation}
\end{figure}

\begin{figure*}[ht]
\centering
\includegraphics[width=0.75\linewidth]{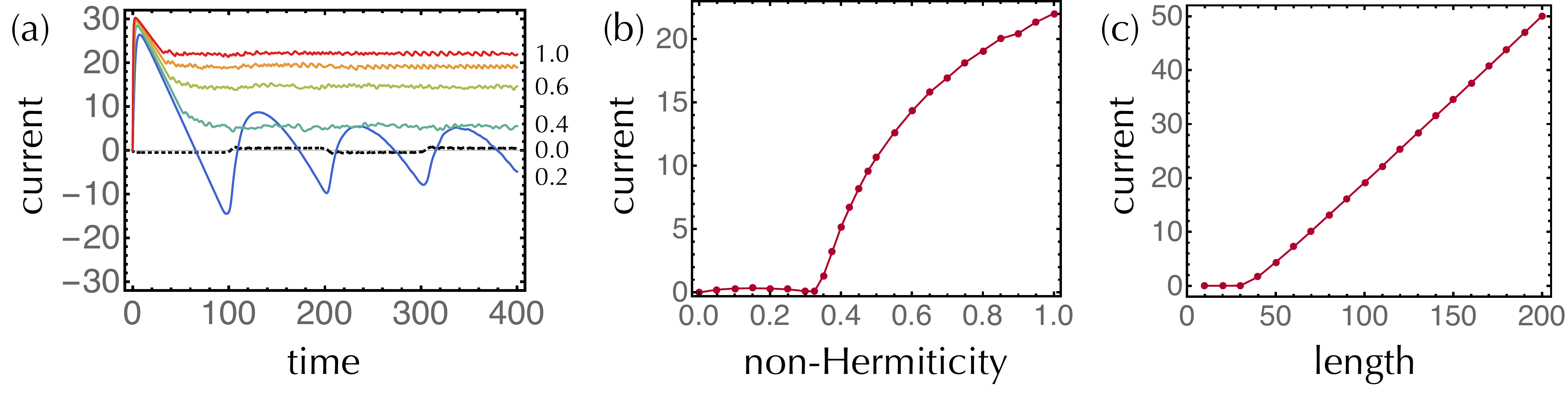} 
\caption{Total charge current $I \left( t \right) \coloneqq \sum_{l=1}^{L-1} \braket{\psi \left( t \right) | \hat{I}_{l} | \psi \left( t \right)}$ in the Hatano-Nelson model with open boundaries ($J=1.0$). 
The initial state is prepared as the charge density wave state in Eq.~(\ref{eq: HN-CDW}).
(a)~Time evolution of the current ($L=100$) for $\gamma = 0.0$ (black dashed curve), $0.2$ (blue curve), $0.4$ (green curve), $0.6$ (light-green curve), $0.8$ (orange curve), and $1.0$ (red curve). 
(b)~Charge current for the steady state as a function of non-Hermiticity $\gamma$ for $L=100$. 
(c)~Charge current for the steady state as a function of the system length for $\gamma = 0.8$.}
	\label{fig: HN-current}
\end{figure*}

We also investigate the time evolution of the correlation matrix
\begin{equation}
    C_{ij} \left( t \right)
    \coloneqq \braket{\psi \left( t \right) | \hat{c}_{i}^{\dag} \hat{c}_{j} | \psi \left( t \right)}
\end{equation}
for $i, j = 1, 2, \cdots, L$.
In the absence of non-Hermiticity, the quasiparticles propagate in both directions, leading to the diffusion of particles and quantum information [Fig.~\ref{fig: HN-correlation}\,(a)].
In the presence of non-Hermiticity, on the other hand, the quasiparticles cease to move, and the correlation propagation is frozen [Fig.~\ref{fig: HN-correlation}\,(b)].
This is another consequence of the skin effect.
Because of the localization of the quasiparticles, they move toward the right (left) edge for $\gamma > 0$ ($\gamma < 0$) at the beginning of the dynamics.
However, once the quasiparticles accumulate at the edge, they are no longer mobile because of the Pauli exclusion principle.
Under the skin effect, the system soon reaches a nonequilibrium steady state in which an extensive number of particles are localized at an edge.
It is noteworthy that the frozen correlation propagation due to the skin effect is different from the supersonic correlation propagation in non-Hermitian quantum systems with reciprocal dissipation~\cite{Ashida-18, Dora-20, *Bacsi-21, Turkeshi-22}. 
This difference also shows a unique role of the skin effect in open quantum systems.

\subsection{Current}

Next, we investigate the charge current
\begin{equation}    
    I_{l} \left( t \right) \coloneqq \braket{\psi \left( t \right) | \hat{I}_{l} | \psi \left( t \right)},
        \label{eq: HN - local current}
\end{equation}
where $\hat{I}_l$ is the local current operator between sites $l$ and $l+1$:
\begin{equation}
    \hat{I}_{l} \coloneqq \frac{\ii J}{2} \left( \hat{c}_{l}^{\dag} \hat{c}_{l+1} - \hat{c}_{l+1}^{\dag} \hat{c}_{l} \right).
\end{equation}
While 
no current flows
in closed quantum systems at thermal equilibrium, the skin effect gives rise to a current in open quantum systems.
Figure~\ref{fig: HN-current} shows the behavior of the total charge current $I \left( t \right) \coloneqq \sum_{l=1}^{L-1} I_{l} \left( t \right)$ induced by the skin effect.
In the presence of non-Hermiticity, the current takes a nonzero steady value for sufficiently long time [Fig.~\ref{fig: HN-current}\,(a)].
This means that the system reaches a nonequilibrium steady state accompanying a nonzero current in contrast with the thermal equilibrium states, where the current should vanish [i.e., $I = o \left( L \right)$]~\cite{Watanabe-19}.
The current for the steady state monotonically increases as a function of non-Hermiticity [Fig.~\ref{fig: HN-current}\,(b)].
Furthermore, it grows linearly with respect to the system length $L$ [Fig.~\ref{fig: HN-current}\,(c)] and hence is indeed a macroscopic quantity.
The macroscopic current induced by the skin effect may be characterized by topological field theory~\cite{KSR-21}.

\begin{figure}[t]
\centering
\includegraphics[width=\linewidth]{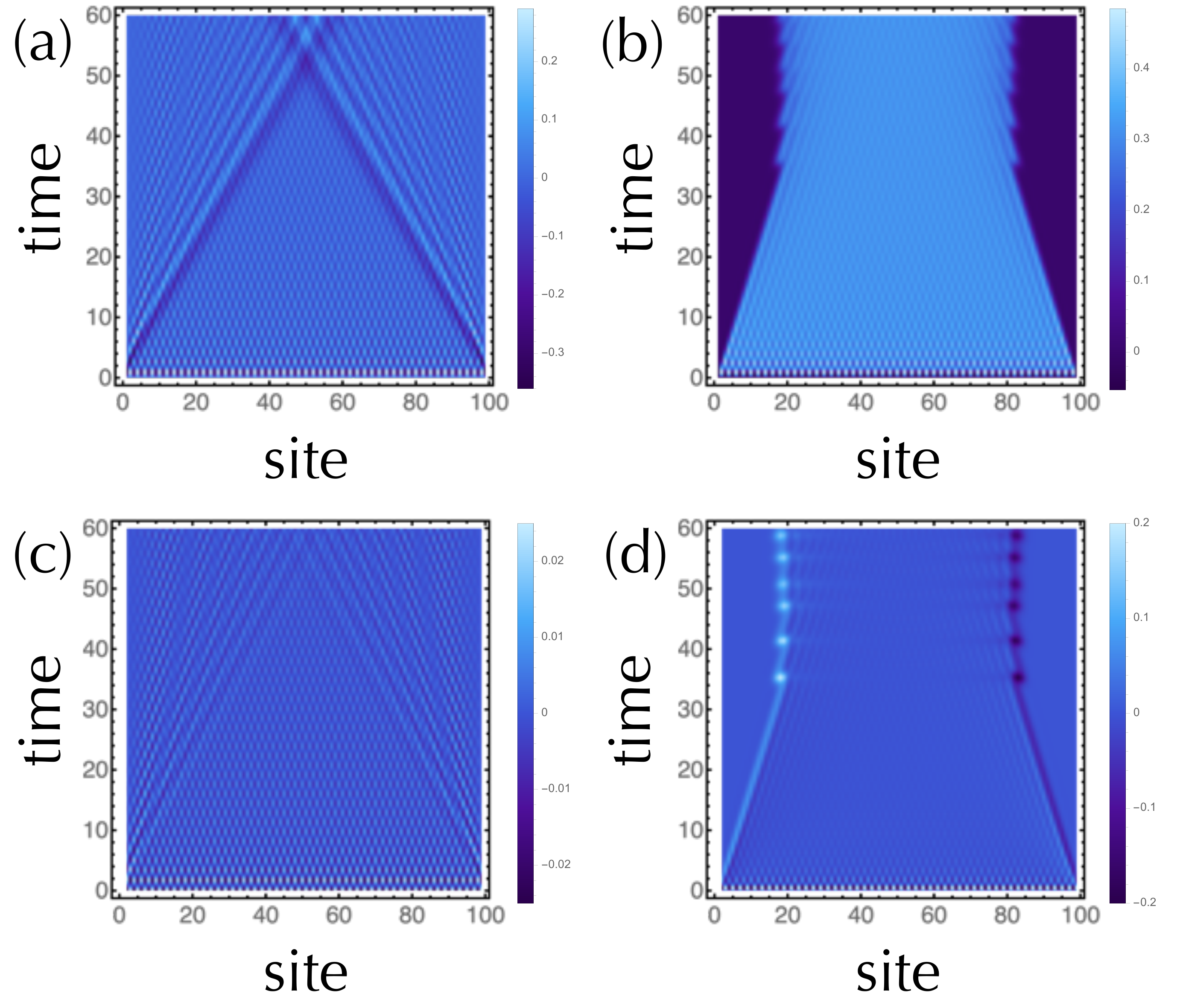} 
\caption{Local current distribution in the Hatano-Nelson model with open boundaries ($J=1.0$). 
The initial state is prepared as the charge density wave state in Eq.~(\ref{eq: HN-CDW}).
(a, b)~Time evolution of the local current $I_{l} \left( t \right) \coloneqq \braket{\psi \left( t \right) | \hat{I}_{l} | \psi \left( t \right)}$ for (a)~$\gamma = 0.0$ and (b)~$\gamma = 0.8$. 
(c, d)~Time evolution of the local particle inflow for (c)~$\gamma = 0.0$ and (d)~$\gamma = 0.8$.}
	\label{fig: HN-current-distribution}
\end{figure}

To understand how the skin effect gives rise to a current in more detail, we also study the local distribution of the current (Fig.~\ref{fig: HN-current-distribution}).
Notably, in the presence of non-Hermiticity, the current arises only in the bulk and vanishes around the edges [Fig.~\ref{fig: HN-current-distribution}\,(b)].
On the basis of the local particle distribution in Fig.~\ref{fig: HN-number}\,(b), the current arises only in the region where the particles are neither dense nor sparse.
This is because particles cannot enter such dense or sparse regions from the environment because of the Pauli exclusion principle. 
It is also compatible with the frozen correlation propagation in Fig.~\ref{fig: HN-correlation}\,(b).
Moreover, the continuity equation of our non-Hermitian system reads
\begin{equation}
    \frac{\partial}{\partial t} n_l + \left( I_{l} - I_{l-1} \right) = \sigma_l,
\end{equation}
where $\sigma_l$ is the local inflow of particles from the external environment at site $l$.
In Hermitian systems, $\sigma_l$ vanishes for arbitrary $l$ and $t$ owing to the conservation of the particle number [Fig.~\ref{fig: HN-current-distribution}\,(c)].
Under the skin effect, a pair of a source and sink appears, between which the current flows [Fig.~\ref{fig: HN-current-distribution}\,(d)].
It is also notable that the current does not arise for small non-Hermiticity or a short system length (Fig.~\ref{fig: HN-current}).
In such a case, the localization length of the many-body skin modes is comparable with the system length, and consequently particles cannot enter the system from the environment.

\subsection{Entanglement dynamics}
    \label{sec: HN-entanglement}

\begin{figure}[t]
\centering
\includegraphics[width=\linewidth]{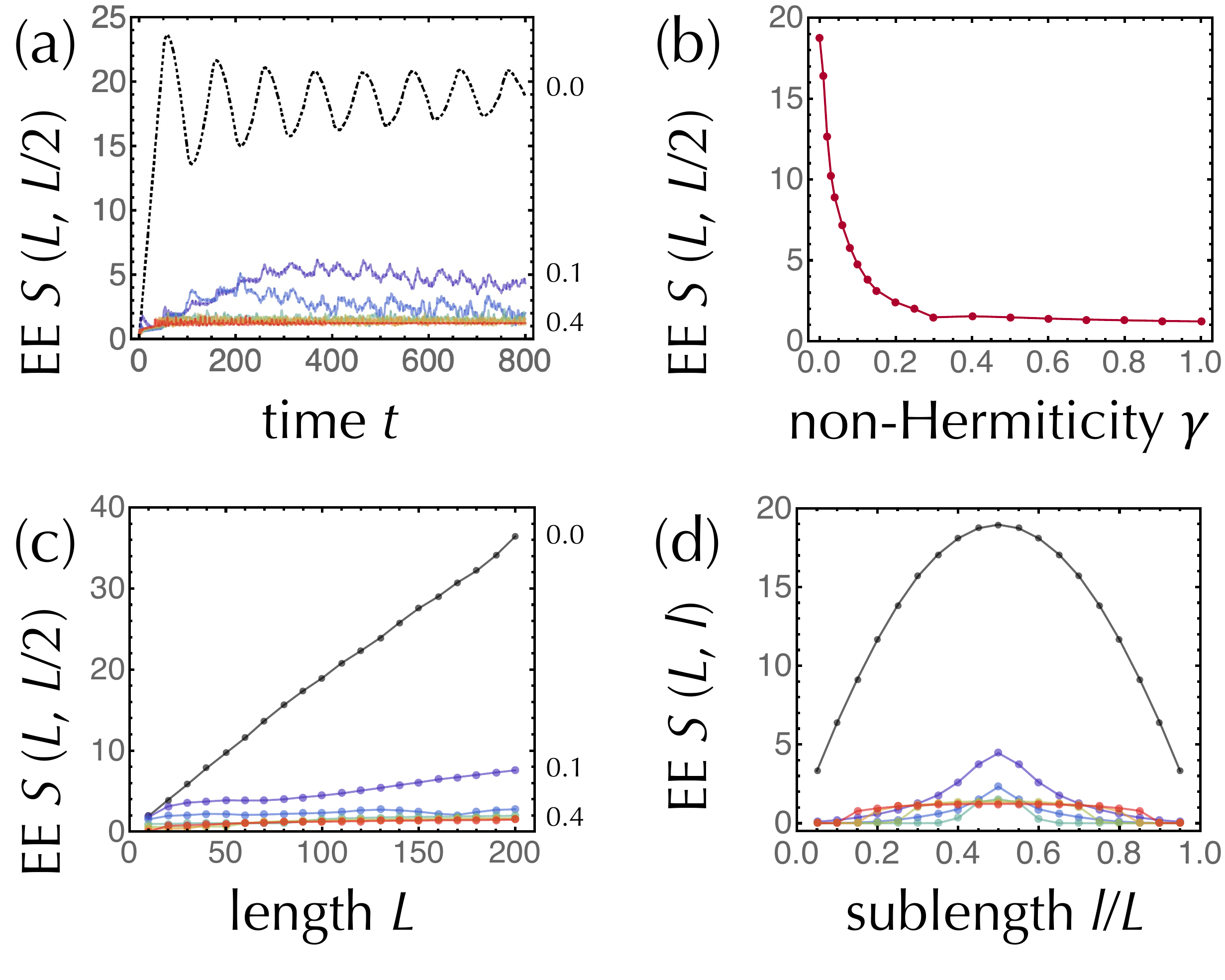} 
\caption{Entanglement entropy of the Hatano-Nelson model with open boundaries ($J=1.0$). 
The initial state is prepared as the charge density wave state in Eq.~(\ref{eq: HN-CDW}).
(a)~Time evolution of the entanglement entropy $S \left( L, L/2 \right)$ ($L=100$) for $\gamma = 0.0$ (black dashed curve), $0.1$ (violet curve), $0.2$ (blue curve), $0.4$ (green curve), $0.6$ (light-green curve), $0.8$ (orange curve), and $1.0$ (red curve). 
(b)~Entanglement entropy $S \left( L, L/2 \right)$ for the steady state as a function of non-Hermiticity $\gamma$ ($L=100$). 
(c)~Entanglement entropy $S \left( L, L/2 \right)$ for the steady state as a function of the system length $L$. (d)~Entanglement entropy $S \left( L, l \right)$ ($L=100$) for the steady state as a function of the subsystem length $l$.}
	\label{fig: HN-EE}
\end{figure}

The non-Hermitian skin effect 
gives rise to a nonequilibrium flow not only of particles but also of quantum information.
To show this, we investigate the time evolution of the entanglement entropy in the Hatano-Nelson model.
We focus on the von Neumann entanglement entropy $S \left( L, l \right)$ between the subsystem in $\left[ 1, l \right]$ and the rest of the system.
Here, we calculate the entanglement entropy from a single wave function $\ket{\psi \left( t \right)}$ instead of the biorthogonal density operator~\cite{Couvreur-17, Herviou-19-ES, Chang-20}.
In Hermitian systems, $S \left( L, l \right)$ grows linearly in time until 
it saturates 
to the extensive entanglement entropy $S \propto l$~\cite{Calabrese-Cardy-05, *Calabrese-Cardy-06, Fagotti-08}, which is consistent with our numerical calculations for $\gamma = 0$ [Fig.~\ref{fig: HN-EE}\,(a)]. 
In the presence of non-Hermiticity, however, the growth of the entanglement entropy is greatly suppressed.
The entanglement entropy for the steady state is much smaller than that for the Hermitian case and monotonically decreases as a function of non-Hermiticity [Fig.~\ref{fig: HN-EE}\,(b)].
In the Hermitian case $\gamma = 0$, the steady-state entanglement entropy grows linearly with the system length, i.e., volume law;
in the non-Hermitian case $\gamma \neq 0$, the steady-state entanglement entropy is independent of the system length, i.e., area law [Fig.~\ref{fig: HN-EE}\,(c, d)].

The suppression of the entanglement entropy originates from the skin effect.
In closed quantum systems, the quasiparticles diffuse throughout the system and let the system be a thermal equilibrium state exhibiting the extensive entanglement entropy.
On the other hand, a macroscopic current from the external environment pushes the quasiparticles only in one direction and forbids quantum diffusion throughout the system.
Consequently, the quasiparticles are localized only at one edge (i.e., skin effect) and cannot develop a global quantum correlation, leading to the area law of the entanglement entropy for the nonequilibrium steady state.

It should be noted that the area law of the entanglement entropy can also occur in non-Hermitian systems with broken parity-time symmetry~\cite{Gopalakrishnan-21}.
In such systems, the suppression of the entanglement is due to the relaxation toward a pure state with the largest imaginary part of the complex-valued energy.
By contrast, our non-Hermitian system hosts the entirely real spectrum under the open boundary conditions and hence does not rely on parity-time-symmetry breaking.
The non-Hermitian skin effect is a new mechanism of open quantum systems that hinders the growth of the quantum correlation and entanglement.

\section{Entanglement phase transition induced by the non-Hermitian skin effect}
    \label{sec: sHN}

In the Hatano-Nelson model, even infinitesimal non-Hermiticity induces the skin effect and makes the system relax to far from equilibrium.
To understand the nonequilibrium quantum criticality induced by the skin effect, we consider the symplectic generalization of the Hatano-Nelson model~\cite{OKSS-20, *KOS-20, KR-21}:
\begin{align}
    \hat{H} &= - \frac{1}{2} \sum_{l=1}^{L} \left[ \hat{c}_{l+1}^{\dag} \left( J+\gamma\sigma_z-\ii\Delta\sigma_x \right)  \hat{c}_{l} \right. \nonumber \\ 
    &\qquad\qquad\qquad \left. + \hat{c}_{l}^{\dag} \left( J-\gamma\sigma_z+\ii\Delta\sigma_x \right) \hat{c}_{l+1} \right]
        \label{eq: sHN}
\end{align}
with Pauli matrices $\sigma_i$'s ($i = x, y, z$).
The fermionic annihilation operator $\hat{c}_{l} = \left( \hat{c}_{l, \uparrow}~\hat{c}_{l, \downarrow} \right)^{T}$ [creation operator $\hat{c}_{l}^{\dag} = ( \hat{c}_{l, \uparrow}^{\dag}~\hat{c}_{l, \downarrow}^{\dag})$] now includes the spin degree of freedom.
Because of non-Hermiticity $\gamma > 0$ ($\gamma < 0$), the up-spin fermions are pushed toward the right (left) while the down-spin fermions are pushed toward the left (right).
In addition, $\Delta \in \mathbb{R}$ controls the spin-orbit coupling between the up-spin fermions and down-spin fermions.
Owing to the spin-orbit coupling $\Delta$, the model is free from the skin effect even in the presence of non-Hermiticity $\gamma$ as long as $\left| \gamma \right| < \left| \Delta \right|$ is satisfied.
Similarly to the original Hatano-Nelson model, the symplectic Hatano-Nelson model in Eq.~(\ref{eq: sHN}) can be implemented in the quantum trajectory approach (see Appendix~\ref{asec: NH} for details).
It is notable that non-Hermitian spin-orbit-coupled fermions have been realized in recent experiments of ultracold atoms~\cite{Ren-22}, 
and our model can also be realized in a similar experiment.

\begin{figure}[t]
\centering
\includegraphics[width=0.55\linewidth]{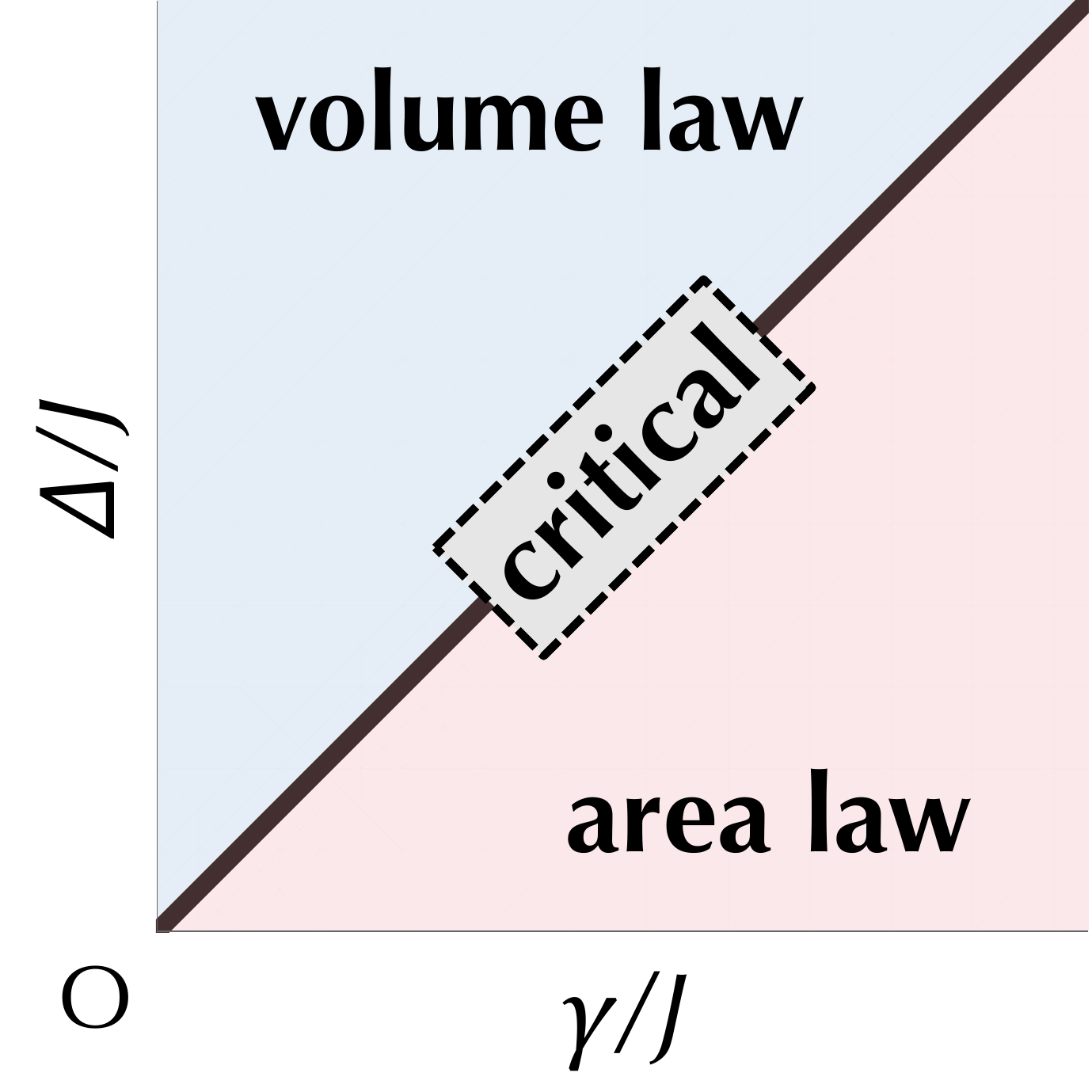} 
\caption{Phase diagram of the symplectic Hatano-Nelson model. For $\left| \gamma \right| < \left| \Delta \right|$ (blue region), no skin effect occurs, and the entanglement entropy for the steady state obeys the volume law. For $\left| \gamma \right| > \left| \Delta \right|$ (red region), the reciprocal skin effect occurs, and the entanglement entropy for the steady state obeys the area law. The phase boundary $\left| \gamma \right| = \left| \Delta \right| \neq 0$ (black line) marks critical points, at which the skin modes exhibit the scale invariance, and the entanglement entropy for the steady state grows subextensively (i.e., logarithmically with respect to the subsystem length).}
	\label{fig: sHN-phase-diagram}
\end{figure}

Under the periodic boundary conditions, the Bloch Hamiltonian of the symplectic Hatano-Nelson model reads
\begin{equation}
    H \left( k \right) 
    = - J \cos k + \left( \ii \gamma \sigma_z + \Delta \sigma_x \right) \sin k,
        \label{eq: sHN-Bloch}
\end{equation}
whose complex spectrum is obtained as
\begin{equation}
    E \left( k \right) = - J \cos k \pm \ii \sqrt{\gamma^2 - \Delta^2} \sin k.
        \label{eq: sHN-spectrum}
\end{equation}
Therefore, for small non-Hermiticity $\left| \gamma \right| < \left| \Delta \right|$, the spectrum is entirely real, and no skin effect occurs.
For large non-Hermiticity $\left| \gamma \right| > \left| \Delta \right|$, on the other hand, each band is characterized by the complex-spectral winding and subject to the skin effect~\cite{OKSS-20, *KOS-20}.
There, up-spin fermions and down-spin fermions are localized at opposite boundaries.
This reciprocal skin effect is ensured by the $\mathbb{Z}_2$ topological invariant 
$\nu \in \{ 0, 1 \}$
unique to non-Hermitian systems~\cite{KSUS-19}:
\begin{align}
    &\left( -1 \right)^{\nu} \coloneqq \mathrm{sgn} \left\{ \frac{\mathrm{Pf} \left[ H \left( k=\pi \right) T \right]}{\mathrm{Pf} \left[ H \left( k=0 \right) T \right]} \right. \nonumber \\
    &\qquad\qquad \left. \times \exp \left[ - \frac{1}{2} \int^{k=\pi}_{k=0} d\log \det \left[ H \left( k \right) T \right] \right] \right\}
        \label{eq: Z2 inv}
\end{align}
with the unitary operator $T \coloneqq \sigma_y$ for the symplectic Hatano-Nelson model.
The presence or absence of the skin effect is controlled by the competition between non-Hermiticity $\gamma$ and spin-orbit coupling $\Delta$, and $\left| \gamma \right| = \left| \Delta \right|$ marks a phase transition point, between which the skin effect occurs or not (Fig.~\ref{fig: sHN-phase-diagram}).
The reciprocal skin effect generally occurs as long as the $\mathbb{Z}_2$ topological invariant in Eq.~(\ref{eq: Z2 inv}) is nontrivial.
Thus, while we here consider the symplectic Hatano-Nelson model in Eq.~(\ref{eq: sHN}) for illustrative purposes, the $\mathbb{Z}_2$ skin effect and the concomitant entanglement phase transition should appear in a wide variety of open quantum systems.

It is also notable that the symplectic Hatano-Nelson model respects reciprocity, which is one of the fundamental internal symmetry for non-Hermitian systems~\cite{KSUS-19}.
In fact, the non-Hermitian Hamiltonian in Eq.~(\ref{eq: sHN}) respects reciprocity
\begin{equation}
    \hat{T} \hat{H}^{\dag} \hat{T}^{-1} = \hat{H},
        \label{eq: sHN-reciprocity}
\end{equation}
where $\hat{T}$ is an antiunitary operator satisfying $\hat{T} \hat{c}_{l} \hat{T}^{-1} = \sigma_y \hat{c}_{l}$ and $\hat{T} z \hat{T}^{-1} = z^{*}$ for $z \in \mathbb{C}$.
In terms of the Bloch Hamiltonian in Eq.~(\ref{eq: sHN-Bloch}), reciprocity is written as
\begin{equation}
    T H^T \left( k \right) T^{-1}
    = H \left( -k \right),\quad
    TT^{*} = -1
        \label{eq: sHN-reciprocity-Bloch}
\end{equation}
with the unitary operator $T \coloneqq \sigma_y$.
The Kramers pair structure between up-spin and down-spin fermions, as well as the concomitant skin effect, is protected by reciprocity.

Below, we study the nonequilibrium quantum dynamics of the symplectic Hatano-Nelson model. 
We choose the initial state as
\begin{equation}
    \ket{\psi_0} = \left( \prod_{l=1}^{L/2} \hat{c}_{2l-1, \uparrow}^{\dag} \hat{c}_{2l, \downarrow}^{\dag} \right) \ket{\rm vac},
        \label{eq: sHN-initial-state}
\end{equation}
where the system length $L$ is assumed to be even.
We confirm that the system reaches a many-body steady state subject to the reciprocal skin effect in Sec.~\ref{sec: sHN - skin}.
This nonequilibrium steady state is characterized by a spin current in contrast to the thermal equilibrium states, as shown in Sec.~\ref{sec: sHN - current}.
Furthermore, in Sec.~\ref{sec: sHN - entanglement}, we demonstrate that the phase boundary $\left| \gamma \right| = \left| \Delta \right|$ marks an entanglement phase transition, between which the steady state exhibits the volume law or the area law (Fig.~\ref{fig: sHN-phase-diagram}).
The critical point $\left| \gamma \right| = \left| \Delta \right|$ is characterized by a conformal field theory that is anomalously sensitive to the boundary conditions.
In Sec.~\ref{sec: sHN-EP}, we also show that this nonequilibrium quantum criticality originates from the scale-invariant skin modes decaying according to the power law.
While we here choose Eq.~(\ref{eq: sHN-initial-state}) as an initial state, the universal properties of the entanglement phase transition---the critical behaviors in Eqs.~(\ref{eq: sHN-c-OBC}), (\ref{eq: sHN - localization length - critical}), and (\ref{eq: sHN - critical skin - lattice})---arise solely from the scale invariance of the skin modes and should not depend on the specific details of the initial state (see Appendix~\ref{asec: initial conditions} for details).

\subsection{Reciprocal skin effect}
    \label{sec: sHN - skin}

\begin{figure*}[t]
\centering
\includegraphics[width=0.9\linewidth]{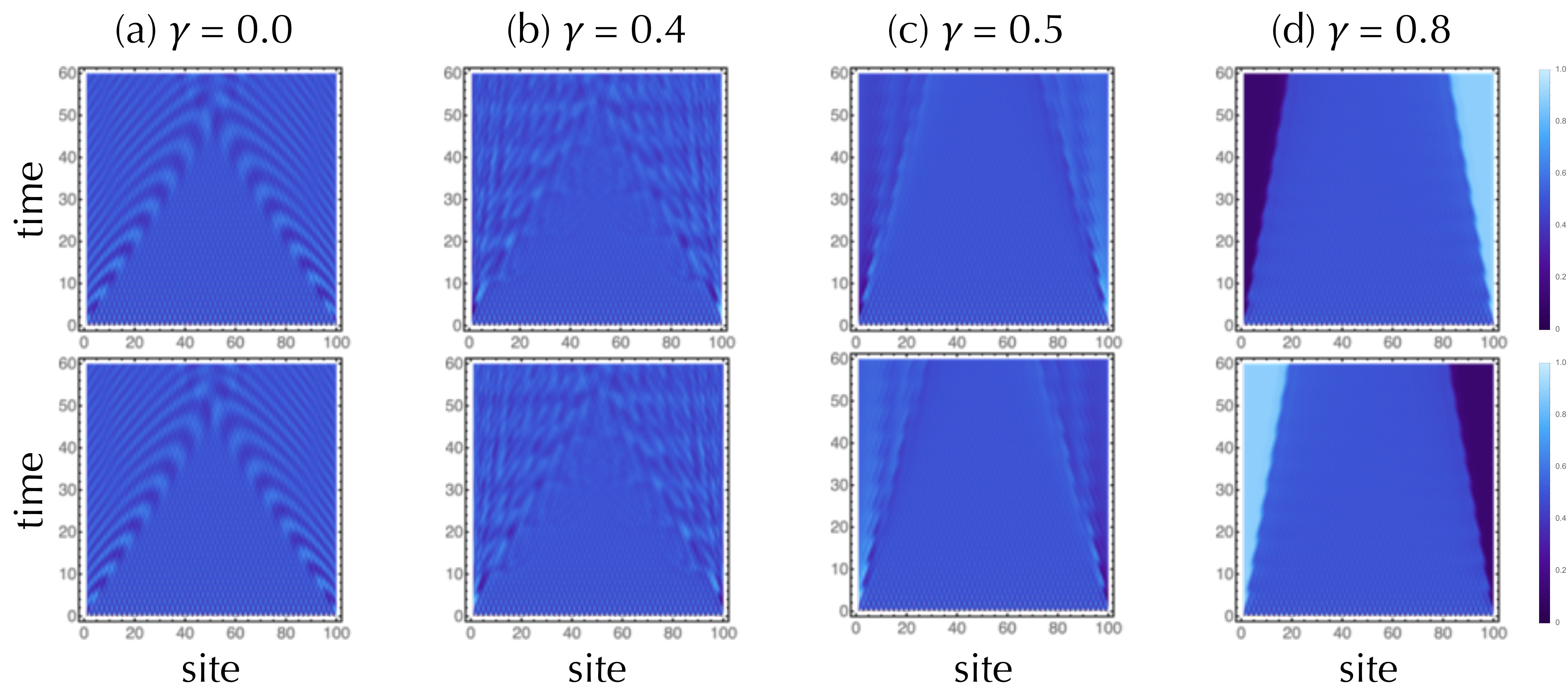} 
\caption{Time evolution of the local particle number $n_{l, s} \left( t \right) \coloneqq \braket{\psi \left( t \right) | \hat{n}_{l, s} | \psi \left( t \right)}$ for $s = \uparrow$ (top panels) and $s=\downarrow$ (bottom panels) in the symplectic Hatano-Nelson model with open boundaries ($L=100$, $J=1.0$, $\Delta = 0.5$). 
The initial state is prepared as Eq.~(\ref{eq: sHN-initial-state}).
Non-Hermiticity is chosen to be (a)~$\gamma = 0.0$, (b)~$\gamma = 0.4$, (c)~$\gamma = 0.5$, and (d)~$\gamma = 0.8$. 
While no skin effect occurs for $\left| \gamma \right| < \left| \Delta \right|$, the reciprocal skin effect occurs for $\left| \gamma \right| > \left| \Delta \right|$.}
	\label{fig: sHN-number}
\end{figure*}

We begin with investigating the time evolution of local particle numbers for each spin (Fig.~\ref{fig: sHN-number}).
Below the critical point (i.e., $\left| \gamma \right| < \left| \Delta \right|$), the particles are distributed almost uniformly throughout the system.
Above the critical point (i.e., $\left| \gamma \right| > \left| \Delta \right|$), on the other hand, the skin effect indeed occurs, and the particles are localized at the edges.
In contrast to the original Hatano-Nelson model, up-spin fermions are localized at the right (left) edge while down-spin fermions are localized at the left (right) edge for $\gamma > 0$ ($\gamma < 0$) [Fig.~\ref{fig: sHN-number}\,(d)].
Consequently, particles are uniformly distributed on average.
This is a unique feature of the reciprocity-protected skin effect in the symplectic Hatano-Nelson model.

\begin{figure*}[t]
\centering
\includegraphics[width=0.9\linewidth]{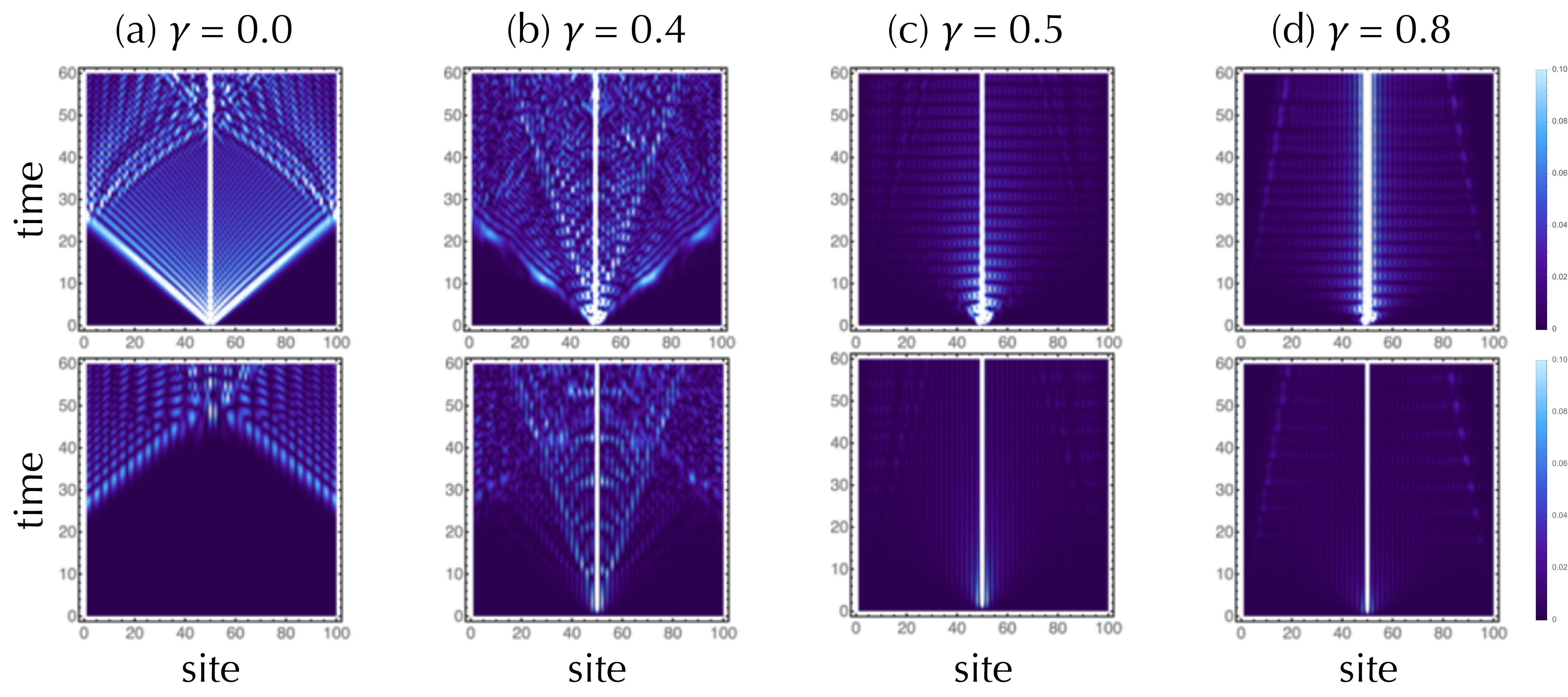} 
\caption{Correlation propagation in the symplectic Hatano-Nelson model with open boundaries ($L=100$, $J=1.0$, $\Delta = 0.5$). 
The absolute values $\left| C_{l\uparrow, l_0\uparrow} \right| = \left| C_{l\downarrow, l_0\downarrow} \right|$ (top panels) and $\left| C_{l\uparrow, l_0\downarrow} \right| = \left| C_{l\downarrow, l_0\uparrow} \right|$ (bottom panels) of the correlation matrix are shown as a function of site $l$ and time $t$ with $l_0 = L/2 = 50$. 
The initial state is prepared as Eq.~(\ref{eq: sHN-initial-state}).
Non-Hermiticity is chosen to be (a)~$\gamma = 0.0$, (b)~$\gamma = 0.4$, (c)~$\gamma = 0.5$, and (d)~$\gamma = 0.8$.}
	\label{fig: sHN-correlation}
\end{figure*}

We also investigate the correlation propagation in the symplectic Hatano-Nelson model (Fig.~\ref{fig: sHN-correlation}).
The correlation matrix now includes the spin degree of freedom:
\begin{equation}
    C_{is, js'} \left( t \right)
    \coloneqq \braket{\psi \left( t \right) | \hat{c}_{i, s}^{\dag} \hat{c}_{j, s'} | \psi \left( t \right)}.
\end{equation}
Below the critical point (i.e., $\left| \gamma \right| < \left| \Delta \right|$), the correlation bidirectionally propagates throughout the system even in the presence of non-Hermiticity, which is a signature of the quantum diffusion.
Above the critical point (i.e., $\left| \gamma \right| > \left| \Delta \right|$), the skin effect freezes the correlation propagation in a similar manner to the original Hatano-Nelson model.
Notably, the quasiparticles cease to propagate even at the critical point (i.e., $\left| \gamma \right| = \left| \Delta \right|$).
The frozen correlation propagation implies the skin effect even at the critical point.
In Sec.~\ref{sec: sHN-EP}, we indeed demonstrate the skin effect at the critical point while the critical skin modes are localized algebraically instead of exponentially.

\subsection{Spin current}
    \label{sec: sHN - current}

\begin{figure}[t]
\centering
\includegraphics[width=\linewidth]{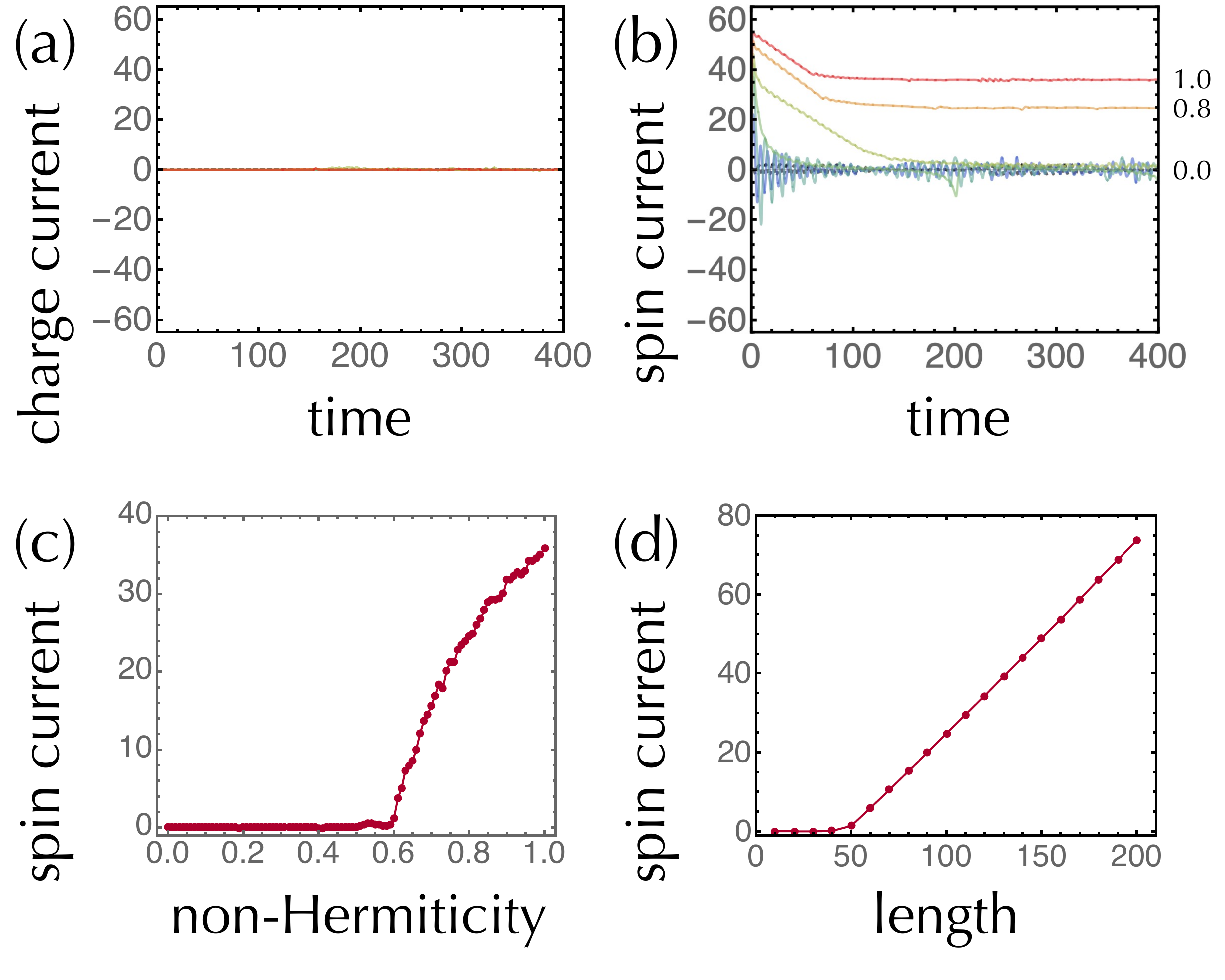} 
\caption{Current in the symplectic Hatano-Nelson model with open boundaries ($J=1.0$, $\Delta = 0.5$). 
The initial state is prepared as Eq.~(\ref{eq: sHN-initial-state}).
(a, b)~Time evolution of the charge current $I_{\rm c} \left( t \right) \coloneqq \braket{\psi \left( t \right) | \hat{I}_{\rm c} | \psi \left( t \right)}$ [(a)] and spin current $I_{\rm s} \left( t \right) \coloneqq \braket{\psi \left( t \right) | \hat{I}_{\rm s} | \psi \left( t \right)}$ [(b)] for $L=100$ and $\gamma = 0.0$, $0.2$, $0.4$, $0.5$, $0.6$, $0.8$, $1.0$. 
(c)~Spin current for the steady state as a function of non-Hermiticity $\gamma$ for $L=100$. 
(d)~Spin current for the steady state as a function of the system length $L$ for $\gamma = 0.8$.}
	\label{fig: sHN-current}
\end{figure}

We next investigate the time evolution of the current.
Owing to the spin degree of freedom, we consider both the total charge current
\begin{equation}
    \hat{I}_{\rm c} \coloneqq \hat{I}_{\uparrow} + \hat{I}_{\downarrow}
\end{equation}
and the total spin current
\begin{equation}
    \hat{I}_{\rm s} \coloneqq \hat{I}_{\uparrow} - \hat{I}_{\downarrow}
\end{equation}
with
\begin{equation}
    \hat{I}_{s} \coloneqq \frac{\ii J}{2} \sum_{l=1}^{L-1} \left( \hat{c}_{l, s}^{\dag} \hat{c}_{l+1, s} - \hat{c}_{l+1, s}^{\dag} \hat{c}_{l, s} \right) \quad \left( s = \uparrow, \downarrow \right).
\end{equation} 
While $\hat{I}_{s}$ is not conserved in the presence of the spin-orbit coupling $\Delta$, it gives an intuitive measure for the spin current.
Even in the presence of non-Hermiticity $\gamma$, the charge current $I_{\rm c} \left( t \right)$ always vanishes as a consequence of reciprocity [Fig.~\ref{fig: sHN-current}\,(a)].
On the other hand, the spin current $I_{\rm s} \left( t \right)$ exhibits characteristic behavior unique to the symplectic Hatano-Nelson model.
Below the critical point (i.e., $\left| \gamma \right| < \left| \Delta \right|$), the spin current just oscillates and vanishes 
after averaging over time;
above the critical point (i.e., $\left| \gamma \right| > \left| \Delta \right|$), the skin effect occurs and induces a nonzero spin current.
Similarly to the steady-state charge current in the original Hatano-Nelson model, the steady-state spin current grows as 
we increase non-Hermiticity or the system length
[Fig.~\ref{fig: sHN-current}\,(c, d)].
Thus, the system reaches a nonequilibrium steady state with a nonzero spin current. 
The spin current characterizes the nonequilibrium quantum phases of the symplectic Hatano-Nelson model as an order parameter.
This is contrasted with the thermal equilibrium states and the nonequilibrium steady states in the original Hatano-Nelson model, which are respectively characterized by zero current and nonzero charge currents.

\subsection{Entanglement phase transition}
    \label{sec: sHN - entanglement}

\begin{figure}[t]
\centering
\includegraphics[width=\linewidth]{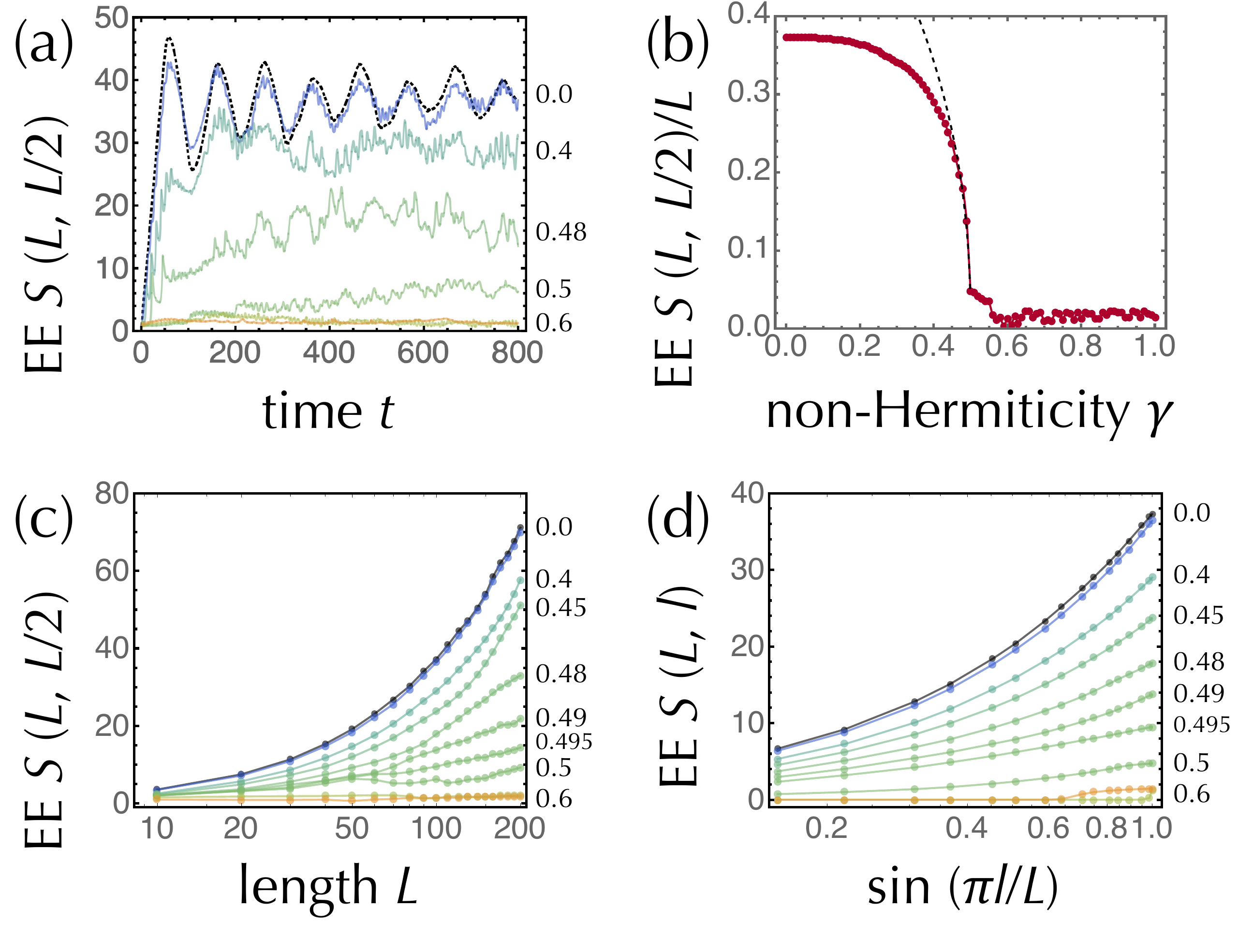} 
\caption{Entanglement entropy of the symplectic Hatano-Nelson model with open boundaries ($J=1.0$, $\Delta = 0.5$). 
The initial state is prepared as Eq.~(\ref{eq: sHN-initial-state}).
(a)~Time evolution of the entanglement entropy $S \left( L, L/2 \right)$ ($L=100$) for $\gamma = 0.0$ (black dashed curve), $0.2$ (blue curve), $0.4$, $0.48$, $0.5$ (green curves), $0.6$ (light-green curve), and $0.8$ (orange curve).
(b)~Entanglement entropy density $S \left( L, L/2 \right)/L$ ($L=100$) for the steady state as a function of non-Hermiticity $\gamma$.
The black dashed curve is the fitting result $S \left( L, L/2 \right)/L = 0.94 \left( \Delta/J - \gamma/J \right)^{0.44}$ around the critical point $\gamma = \Delta$.
(c)~Entanglement entropy $S \left( L, L/2 \right)$ for the steady state as a function of the system length $L$ for $\gamma = 0.0$, $0.2$, $0.4$, $0.45$, $0.48$, $0.49$, $0.495$, $0.5$, $0.6$, and $0.8$. 
(d)~Entanglement entropy $S \left( L, l \right)$ ($L=100$) for the steady state as a function of the subsystem length $l$.}
	\label{fig: sHN-EE}
\end{figure}

Now, we investigate the entanglement dynamics of the symplectic Hatano-Nelson model (Fig.~\ref{fig: sHN-EE}).
In the Hermitian case $\gamma = 0$, 
the system reaches the thermal equilibrium state (or the generalized Gibbs state) under the dynamics, and
the entanglement entropy for the steady state grows linearly with the system length, i.e., volume law.
Even in the presence of non-Hermiticity, the volume law of the entanglement entropy persists for $\left| \gamma \right| < \left| \Delta \right|$. 
This contrasts with the original Hatano-Nelson model, in which the volume law is violated by infinitesimal non-Hermiticity (Sec.~\ref{sec: HN-entanglement}).
The robust volume law is consistent with the quantum diffusion of quasiparticles shown in Fig.~\ref{fig: sHN-correlation}.
As non-Hermiticity increases, the entanglement entropy for the steady state gradually decreases and sharply vanishes at $\left| \gamma \right| = \left| \Delta \right|$.
For the larger non-Hermiticity $\left| \gamma \right| > \left| \Delta \right|$, the entanglement entropy is greatly suppressed and no longer grows even if we increase the system length $L$, i.e., the area law.
Similarly to the original Hatano-Nelson model, the area law of the steady-state entanglement entropy arises from the skin effect.
Here, $\left|\gamma \right| = \left| \Delta\right|$ marks a nonequilibrium phase transition across which the steady-state entanglement entropy exhibits the volume law or the area law (Fig.~\ref{fig: sHN-phase-diagram}).
Around this transition point $\left|\gamma \right| = \left| \Delta\right|$, the density of the steady-state entanglement entropy exhibits the critical behavior
\begin{align}
    \frac{S_{\rm s} \left( L, L/2 \right)}{L} \propto \left( \frac{\left| \Delta\right| - \left| \gamma \right|}{J} \right)^{0.44 \pm 0.06}
        \label{eq: sHN-OBC-volume-critical}
\end{align}
for $\left| \gamma \right| \leq \left| \Delta \right|$ [Fig.~\ref{fig: sHN-EE}\,(b)].

Notably, the entanglement phase transition induced by the skin effect occurs even without randomness.
This contrasts with the phase transitions induced by quantum measurements, which typically rely on spatial or temporal randomness~\cite{Chan-19, Skinner-19, Li-18, *Li-19, Choi-20, *Bao-20, Gullans-20, Cao-19, Jian-20, Lavasani-21, Sang-21, Ippoliti-21, Fuji-20, Alberton-21, Ippoliti-21-spacetimeL, *Ippoliti-21-spacetimeX, Lu-21}, although some models can exhibit the phase transitions even without randomness~\cite{Li-19}.
The skin effect provides a new mechanism for the entanglement phase transition and gives rise to a new universality class of nonequilibrium quantum phase transitions.

\begin{figure}[t]
\centering
\includegraphics[width=\linewidth]{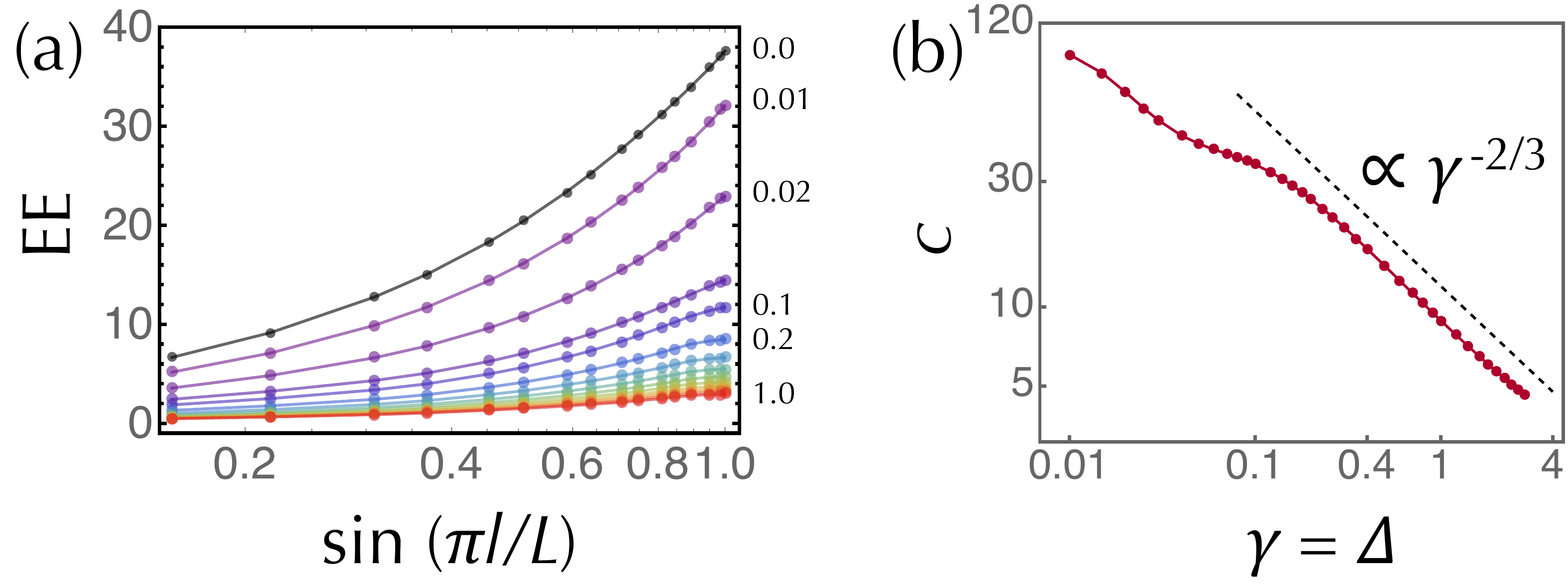} 
\caption{Entanglement entropy of the symplectic Hatano-Nelson model with open boundaries ($J=1.0$) at the critical point ($\gamma = \Delta$). 
The initial state is prepared as Eq.~(\ref{eq: sHN-initial-state}).
(a)~Entanglement entropy $S \left( L, l \right)$ ($L=100$) for the steady state as a function of the subsystem length $l$ for $\gamma = 0.0$, $0.01$, $0.02$, $0.05$, $0.1$, $0.2$, $0.3$, $0.4$, $0.5$, $0.6$, $0.7$, $0.8$, $0.9$, and $1.0$.
(b)~Effective central charge $c$ as a function of $\gamma = \Delta$ ($L=100$).}
	\label{fig: sHN-EE-critical}
\end{figure}

To unveil the nonequilibrium quantum criticality, we further study the entanglement entropy at the transition point $\left| \gamma \right| = \left| \Delta \right|$.
We numerically calculate the steady-state entanglement entropy as a function of the system parameter $\left| \gamma/J \right| = \left| \Delta/J \right|$.
According to the conformal field theory description~\cite{Calabrese-Cardy-04, Calabrese-Cardy-05, *Calabrese-Cardy-06}, the entanglement entropy $S_{\rm s} \left( L, l \right)$ of a one-dimensional quantum critical system with open boundaries grows logarithmically with respect to the subsystem length $l$: 
\begin{equation}
    S_{\rm s} \left( L, l \right) = \frac{c}{6} \log \left( \sin \frac{\pi l}{L} \right) + S_0, 
\end{equation}
where $c$ is the central charge that characterizes the relevant conformal field theory, and $S_0$ is a nonuniversal constant.
Despite non-Hermiticity, the steady-state entanglement entropy of the symplectic Hatano-Nelson model at the critical point $\left| \gamma \right| = \left| \Delta \right|$ is well fitted by this subextensive scaling [Fig.~\ref{fig: sHN-EE-critical}\,(a)].
Remarkably, the effective central charge $c$ is sensitive to the system parameter $\gamma/J = \Delta/J$ in contrast to unitary conformal field theory for closed quantum systems [Fig.~\ref{fig: sHN-EE-critical}\,(b)].
It can take large values for small non-Hermiticity $\left| \gamma/J \right|$, in which a crossover between the unitary and nonunitary critical points should occur.
For larger $\left| \gamma/J \right|$, on the other hand, the effective central charge $c$ exhibits the power-law behavior:
\begin{equation}
    c \propto \left| \gamma/J \right|^{-\left( 0.66 \pm 0.03\right)},
        \label{eq: sHN-c-OBC}
\end{equation}
whose critical exponent is close to $2/3$.
Here, we identify the effective central charge from the logarithmic scaling of the entanglement entropy.
We should note that this is apparently different from the effective central charge in the context of nonunitary conformal field theory, which is defined by subtracting the dimension of the lowest-dimensional operator from the central charge.
Still, the parameter-dependent effective central charge $c$ implies nonunitary or irrational conformal field theory that underlies the nonequilibrium quantum criticality induced by the skin effect.
It merits further study to identify this anomalous type of conformal field theory.

It should also be noted that a couple of recent works on random nonunitary quantum dynamics have reported a similar subextensive growth of the steady-state entanglement entropy with the parameter-dependent effective central charge~\cite{Ippoliti-21, Chen-20, Alberton-21}.
For example, in the nonunitary random dynamics of free fermions in Ref.~\cite{Chen-20}, the effective central charge obeys $c \propto \beta^{-1}$, where $\beta$ is the degree of non-Hermiticity.
The different exponents, $2/3$ of our symplectic Hatano-Nelson model and $1$ of the nonunitary random dynamics in Ref.~\cite{Chen-20},
signal the different universality classes of the entanglement phase transition.
Furthermore, as also discussed above, temporal randomness plays a crucial role in the entanglement phase transitions in Refs.~\cite{Ippoliti-21, Chen-20, Alberton-21}.
By contrast, the entanglement phase transition in this work is based not on the randomness but on the skin effect.
As shown below, it arises from the scale invariance of skin modes, and consequently, the underlying nonunitary conformal field theory is also anomalously sensitive to the boundary conditions.
Our model provides a new type of nonequilibrium quantum phase transitions that belongs to a different universality class.

\subsection{Criticality of skin modes}
    \label{sec: sHN-EP}

We demonstrate that the nonequilibrium quantum criticality at the phase transition point $\left| \gamma \right| = \left| \Delta \right|$ originates from the scale  
invariance of the skin modes due to an exceptional point.
To understand this, we first perform an imaginary gauge transformation in a manner similar to the original Hatano-Nelson model (Sec.~\ref{sec: HN-skin}).
Here, because of the spin degree of freedom, we consider the following $\mathrm{SL} \left( 2 \right)$ gauge transformation rather than the $\mathrm{GL} \left( 1 \right)$ one~\cite{KR-21}:
\begin{equation}
    \hat{p}_{l}^{\dag} \coloneqq \hat{c}_{l}^{\dag} V \begin{pmatrix}
    e^{l\theta} & 0 \\
    0 & e^{-l\theta}
    \end{pmatrix},\quad
    \hat{q}_{l} \coloneqq \begin{pmatrix}
    e^{-l\theta} & 0 \\
    0 & e^{l\theta}
    \end{pmatrix} V^{-1} \hat{c}_{l}
\end{equation}
for $\theta \in \mathbb{C}$ and $V \in \mathrm{SL} \left( 2 \right)$ ($\mathrm{SL} \left( n \right)$ is the special linear group of $n \times n$ matrices with determinant $1$).
This transformation retains reciprocity in Eqs.~(\ref{eq: sHN-reciprocity}) and (\ref{eq: sHN-reciprocity-Bloch}).
With these new fermion operators $\hat{p}_{l}^{\dag}$ and $\hat{q}_{l}$, the symplectic Hatano-Nelson model reads
\onecolumngrid
\begin{align}
    \hat{H} &= - \frac{1}{2} \sum_{l=1}^{L} \left[ \hat{p}_{l+1}^{\dag} \begin{pmatrix} 
    e^{- \left( l+1 \right) \theta} & 0 \\
    0 & e^{\left( l+1 \right) \theta}
    \end{pmatrix}
    V^{-1} \left( J+\gamma\sigma_z-\ii\Delta\sigma_x \right) V \begin{pmatrix} 
    e^{l \theta} & 0 \\
    0 & e^{-l \theta}
    \end{pmatrix} \hat{q}_{l} \right. \nonumber \\ 
    &\qquad\qquad\qquad\qquad\qquad \left. + \hat{p}_{l}^{\dag} \begin{pmatrix} 
    e^{-l \theta} & 0 \\
    0 & e^{l \theta}
    \end{pmatrix} V^{-1} \left( J-\gamma\sigma_z+\ii\Delta\sigma_x \right) V \begin{pmatrix} 
    e^{\left( l+1 \right) \theta} & 0 \\
    0 & e^{- \left( l+1 \right) \theta}
    \end{pmatrix} \hat{q}_{l+1} \right].
\end{align}
\twocolumngrid
\noindent
Away from the critical point $\left| \gamma \right| = \left| \Delta \right|$, the non-Hermitian matrix $J-\gamma\sigma_z+\ii\Delta\sigma_x$ can be diagonalized by appropriately choosing $V$:
\begin{align}
    &V^{-1} \left( J-\gamma\sigma_z+\ii\Delta\sigma_x \right) V \nonumber \\
    &\qquad = \begin{pmatrix}
    J + \sqrt{\gamma^2 - \Delta^2} & 0 \\
    0 & J - \sqrt{\gamma^2 - \Delta^2}
    \end{pmatrix}.
        \label{eq: sHN-diagonalization}
\end{align}
Furthermore, let us choose $\theta$ such that it satisfies $e^{-\theta}\,( J + \sqrt{\gamma^2 - \Delta^2} ) = e^{\theta}\,( J - \sqrt{\gamma^2 - \Delta^2} )$, i.e.,
\begin{equation}
    \theta = \frac{1}{2} \log \left( \frac{J + \sqrt{\gamma^2 - \Delta^2}}{J - \sqrt{\gamma^2 - \Delta^2}} \right).
        \label{eq: sHN-theta}
\end{equation}
With these choices of $V$ and $\theta$, the Hamiltonian reduces to
\begin{equation}
    \hat{H} = - \frac{\sqrt{J^2 - \gamma^2 + \Delta^2}}{2} \sum_{l=1}^{L-1} \left( \hat{p}_{l+1}^{\dag} \hat{q}_{l} + \hat{p}_{l}^{\dag} \hat{q}_{l+1} \right),
\end{equation}
in which the asymmetric hopping vanishes formally.
It can be further diagonalized similarly to Eq.~(\ref{eq: HN-OBC-diagonalization}).
The imaginary gauge transformation is feasible only under the open boundary conditions in such a manner that the boundary conditions are respected.
The spectrum does not show any singular behavior even across the critical point $\left| \gamma \right| = \left| \Delta \right|$, which contrasts with the emergence of an exceptional point under the periodic boundary conditions (see Sec.~\ref{sec: sHN - PBC} for details).

If the skin effect occurs, the localization properties of the skin modes are captured by the quasiparticles $\hat{p}_{l}$ and $\hat{q}_{l}$.
For $\mathrm{Re}\,\theta > 0$, the up-spin (down-spin) component of $\hat{p}_{l}$ is exponentially localized at the right (left) edge while the up-spin (down-spin) component of $\hat{q}_{l}$ is exponentially localized at the left (right) edge.
Here, all the quasiparticles are subject to the skin effect, and no delocalized modes are present in the bulk.
The localization length $\xi$ of the single-particle skin modes is obtained from Eq.~(\ref{eq: sHN-theta}) as
\begin{equation}
    \xi = \frac{1}{\mathrm{Re}\,\theta}
    = \begin{cases}
    \infty & \left( \left| \gamma \right| < \left| \Delta \right| \right); \\
    1/\left| \theta \right| & \left( \left| \gamma \right| > \left| \Delta \right| \right).
    \end{cases}
\end{equation}
Thus, no skin effect occurs for $\left| \gamma \right| < \left| \Delta \right|$ while the reciprocal skin effect occurs for $\left| \gamma \right| > \left| \Delta \right|$, which is consistent with our numerical calculations in Fig.~\ref{fig: sHN-number}.
Notably, around the critical point $\left| \gamma \right| = \left| \Delta \right|$, the localization length $\xi$ exhibits the critical behavior:
\begin{equation}
    \xi \simeq \frac{J}{\sqrt{\gamma^2 - \Delta^2}}
    \propto \left( \left| \gamma \right| - \left| \Delta \right| \right)^{-1/2}.
        \label{eq: sHN - localization length - critical}
\end{equation}

At the critical point $\left| \gamma \right| = \left| \Delta \right|$, the localization length $\xi$ of the skin modes diverges, which signals the scale invariance.
Consequently, we find that there emerge skin modes decaying according to the power law due to an exceptional point.
At the critical point $\left| \gamma \right| = \left| \Delta \right|$, the above imaginary gauge transformation is no longer applicable.
In fact, the non-Hermitian matrix $J - \gamma \sigma_z + \ii \Delta \sigma_x$ is nondiagonalizable for $\left| \gamma \right| = \left| \Delta \right|$ and supports an exceptional point.
Instead of the diagonalization in Eq.~(\ref{eq: sHN-diagonalization}), the matrix is only transformed into the Jordan normal form
\begin{align}
    \left. V^{-1} \left( J-\gamma\sigma_z+\ii\Delta\sigma_x \right) V \right|_{\left| \gamma \right| = \left| \Delta \right|} 
     = \begin{pmatrix}
    J & -\gamma \\
    0 & J
    \end{pmatrix}.
\end{align}
As a result, the Hamiltonian reduces to
\begin{equation}
    \hat{H} = - \frac{J}{2} \sum_{l=1}^{L-1} \left[ \hat{p}_{l+1}^{\dag} \begin{pmatrix}
    1 & \gamma/J \\
    0 & 1
    \end{pmatrix}
    \hat{q}_{l} + \hat{p}_{l}^{\dag} \begin{pmatrix}
    1 & - \gamma/J \\
    0 & 1
    \end{pmatrix} \hat{q}_{l+1} \right].
\end{equation}

Because of the nondiagonalizability, this defective Hamiltonian supports scale-invariant skin modes linearly localized at the boundary. 
To see this, we study the spatial distribution of the single-particle wave functions in a transfer-matrix method (see, for example, Ref.~\cite{Ohtsuki-review}).
Let $E \in \mathbb{C}$ be 
a
single-particle eigenenergy and $\ket{\phi} = \sum_{l, s} \phi_{l,s} \ket{l} \ket{s}$ be the corresponding eigenstate, where $l$ and $s$ denote the sites and spins, respectively.
The single-particle Schr\"odinger equation in real space reads
\begin{equation}
 - \frac{J}{2} \begin{pmatrix}
 1 & \gamma/J \\
 0 & 1
 \end{pmatrix} \vec{\phi}_{l-1} 
 - \frac{J}{2} \begin{pmatrix}
 1 & - \gamma/J \\
 0 & 1
 \end{pmatrix} \vec{\phi}_{l+1}
 = E \vec{\phi}_{l}
\end{equation}
with $\vec{\phi}_{l} = \left( \phi_{l, \uparrow}~\phi_{l, \downarrow} \right)^{T}$.
For simplicity, we consider a zero-energy eigenstate (i.e., $E=0$).
Then, we have
\begin{equation}
    \vec{\phi}_{l+1} = - \begin{pmatrix}
 1 & \gamma/J \\
 0 & 1
 \end{pmatrix}^2 \vec{\phi}_{l-1},
\end{equation}
leading to
\begin{align}
    \vec{\phi}_{2l+1} &= \left( -1 \right)^{l} \begin{pmatrix}
 1 & \gamma/J \\
 0 & 1
 \end{pmatrix}^{2l} \vec{\phi}_{1}, \\
 \vec{\phi}_{2l+2} &= \left( -1 \right)^{l} \begin{pmatrix}
 1 & \gamma/J \\
 0 & 1
 \end{pmatrix}^{2l} \vec{\phi}_{2}.
\end{align}
As an important property of the Jordan normal form, it is nilpotent with index 2, i.e.,  
\begin{equation}
    \left[ \begin{pmatrix}
 1 & \gamma/J \\
 0 & 1
 \end{pmatrix} - 1 \right]^n = 0
\end{equation}
for $n \geq 2$.
Consequently, we have
\begin{equation}
    \| \vec{\phi}_{2l+1} \|
    = \left\| \begin{pmatrix}
 1 & 2l \gamma/J \\
 0 & 1
 \end{pmatrix} \vec{\phi}_{1} \right\|
 \propto \frac{2l \left| \gamma \right|}{J} \| \vec{\phi}_{1} \|
    \label{eq: sHN - critical skin - lattice}
\end{equation}
for sufficiently large $l$, meaning the linear growth of the norm $\| \phi_l \|$ of the wave function with respect to the site $l$.
Thus, the skin modes at the critical point are localized linearly in contrast to the exponentially-localized skin modes off the critical point.
The linear localization of the critical skin modes gets stronger for larger non-Hermiticity $\left| \gamma \right|$, which is compatible with the decrease of entanglement entropy as a function of $\left| \gamma \right|$ (Fig.~\ref{fig: sHN-EE-critical}).
We note that similar power-law decay arises even for $E \neq 0$ since the $l$th power of the Jordan normal form still appears.
It is also notable that the $l$th power of a diagonalizable matrix gives $\lambda^{l}$ with the eigenvalue $\lambda$ of the matrix.
The emergence of the power law in terms of $l$, rather than the exponential, is a unique feature of nondiagonalizable matrices.
In general, the $\left( n-1 \right)$\,th-power-law localization $\| \vec{\phi}_{l} \| \propto l^{-\left( n-1 \right)}$ ($l \gg 1$) appears if an $n \times n$ Jordan matrix is concerned while only the linear localization appears in the symplectic Hatano-Nelson model.

The criticality of skin modes is understood also by a continuum model. 
To have such a continuum model, let us focus on a gapless point $k= \pi/2$, around which the Bloch Hamiltonian $H \left( k \right)$ in Eq.~(\ref{eq: sHN-Bloch}) reads
\begin{equation}
    H \left( k \right) \simeq Jk + \ii \gamma \sigma_z + \Delta \sigma_x.
\end{equation}
Now, we consider a semi-infinite system with a domain wall at $x=0$.
The system is prepared as a vacuum for $x < 0$ while the Hamiltonian for $x > 0$ is
\begin{equation}
    H \left( x \right) = - \ii J \partial_x + \ii \gamma \sigma_z + \Delta \sigma_x.
\end{equation}
Let $E \in \mathbb{R}$ be 
an
eigenenergy 
and $\vec{\phi} \left( x \right) \in \mathbb{C}^2$ be the corresponding right eigenstate.
For $x > 0$, the Schr\"odinger equation reads
\begin{equation}
    \left( - \ii J \partial_x + \ii \gamma \sigma_z + \Delta \sigma_x \right) \vec{\phi} \left( x \right) = E \vec{\phi} \left( x \right),
\end{equation}
which is solved as
\begin{align}
    &\vec{\phi} \left( x \right)
    = e^{\ii \left( E - \ii \gamma \sigma_z + \Delta \sigma_x \right) x/J} \vec{\phi} \left( 0 \right) \nonumber \\
    &~= e^{\ii Ex/J} \left[ \cosh \left( \frac{x}{\xi} \right) + \frac{\left( \gamma \sigma_z + \ii \Delta \sigma_x \right) \xi}{J} \sinh \left( \frac{x}{\xi} \right) \right] \vec{\phi} \left( 0 \right)
\end{align}
with $\xi \coloneqq J/\sqrt{\gamma^2-\Delta^2}$ [i.e., Eq.~(\ref{eq: sHN - localization length - critical})].
Thus, away from the critical point (i.e., $\left| \gamma \right| \neq \left| \Delta \right|$), the wave function for large $x$ behaves as
\begin{equation}
    \| \vec{\phi} \left( x \right) \| \simeq \begin{cases}
    \| \vec{\phi} \left( 0 \right) \| & \left( \left| \gamma \right| < \left| \Delta \right| \right); \\
    e^{x/\xi} \| \vec{\phi} \left( 0 \right) \| & \left( \left| \gamma \right| > \left| \Delta \right| \right),
    \end{cases}
\end{equation}
which is consistent with the results for the corresponding lattice model.
At the critical point $\left| \gamma \right| \neq \left| \Delta \right|$, by contrast, the relevant length scale $\xi$ diverges, and the wave function behaves as
\begin{equation}
    \| \vec{\phi} \left( x \right) \| \simeq \frac{\left| \gamma \right| x}{J} \| \vec{\phi} \left( 0 \right) \|,
\end{equation}
which also reproduces the linear localization of the skin modes [i.e., Eq.~(\ref{eq: sHN - critical skin - lattice})].

The scale invariance at the critical point appears also for thermal phase transitions~\cite{Goldenfeld-textbook, Cardy-textbook} and quantum phase transitions~\cite{Sachdev-textbook} 
in equilibrium.
At such a critical point  
in equilibrium,
the correlation length diverges and the power-law correlation arises.
By contrast, the scale invariance of our non-Hermitian system originates from the exceptional point and the concomitant scale-invariant skin modes, which are intrinsic to open quantum systems.
Our results provide a new type of nonequilibrium quantum criticality that has no analogs in closed quantum systems. 

We note in passing that the phase transition in the symplectic Hatano-Nelson model is distinct from a discontinuous phase transition in Refs.~\cite{Li-20, Yokomizo-21}, which studied the finite-size scaling of skin modes in the presence of a symmetry-breaking perturbation.
In these previous works, skin modes are localized exponentially even at the phase transition point.
By contrast, the symplectic Hatano-Nelson model 
exhibits a continuous phase transition that
hosts skin modes localized according to the power law,
for which the universal critical exponents such as Eqs.~(\ref{eq: sHN-c-OBC}), (\ref{eq: sHN - localization length - critical}), and (\ref{eq: sHN - critical skin - lattice}) are well defined.

\subsection{Criticality for the periodic boundary conditions}
    \label{sec: sHN - PBC}

To understand the significance of the skin effect, we also study the entanglement dynamics of the symplectic Hatano-Nelson model with periodic boundaries.
Under the periodic boundary conditions, the model exhibits a phase transition also at $\left| \gamma \right| = \left| \Delta \right|$.
However, the phase transition is not characterized by the skin effect but the reality of the spectrum.
In fact, eigenstates are always delocalized throughout the system because of translation invariance.
Meanwhile, the spectrum $E \left( k \right)$ in Eq.~(\ref{eq: sHN-spectrum}) is entirely real for $\left| \gamma \right| \leq \left| \Delta \right|$ but no longer real for $\left| \gamma \right| > \left| \Delta \right|$.
At the critical point $\left| \gamma \right| = \left| \Delta \right|$, the Bloch Hamiltonian $H \left( k \right)$ in Eq.~(\ref{eq: sHN-Bloch}) is not diagonalizable and forms an exceptional point.

\begin{figure}[t]
\centering
\includegraphics[width=\linewidth]{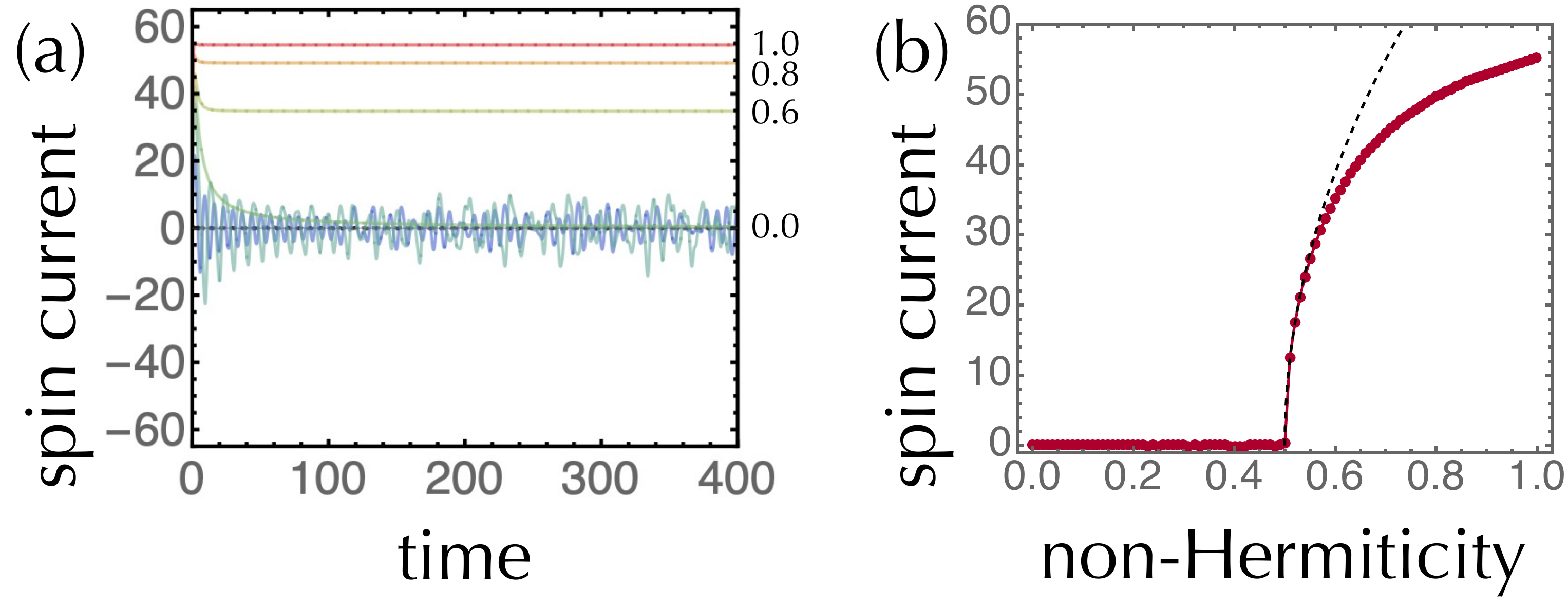} 
\caption{Spin current in the symplectic Hatano-Nelson model with periodic boundaries ($L=100$, $J=1.0$, $\Delta = 0.5$).
The initial state is prepared as Eq.~(\ref{eq: sHN-initial-state}).
(a)~Time evolution of the spin current for $\gamma = 0.0$ (black dashed curve), $0.2$ (blue curve), $0.4$, $0.5$, $0.6$ (green curves), $0.8$ (orange curve), and $1.0$ (red curve). 
(b)~Spin current for the steady state as a function of non-Hermiticity $\gamma$.
The black dashed curve is the fitting result $I_{\rm s} = 123 J \left( \gamma/J - \Delta/J \right)^{0.50}$ around the critical point $\gamma = \Delta$.
}
	\label{fig: sHN-PBC-current}
\end{figure}

Similarly to the open boundary conditions, the time-averaged spin current vanishes below the critical point (Fig.~\ref{fig: sHN-PBC-current}).
Above the critical point, the spectrum is complex, and the system relaxes to the many-body eigenstate that possesses the largest imaginary part of the complex spectrum.
This nonequilibrium steady state is characterized by the nonzero spin current, which is qualitatively similar to the spin current induced by the skin effect (Fig.~\ref{fig: sHN-current}).
It should be noted that the spin current for the open boundary conditions is carried by a superposition of many-body skin modes instead of a single eigenstate.
Around the critical point, the steady-state spin current exhibits the power-law behavior
\begin{align}
    I_{\rm s} \propto J \left( \frac{\left| \gamma \right| - \left| \Delta \right|}{J} \right)^{0.50 \pm 0.02}\quad \left( \left| \gamma \right| \geq \left| \Delta \right| \right),
\end{align}
where the critical exponent $0.50 \pm 0.02$ is close to $1/2$.
This critical exponent may be related to the point-gap closing and the concomitant emergence of an exceptional point, where the complex spectrum in Eq.~(\ref{eq: sHN-spectrum}) exhibits the similar critical behavior $\mathrm{Im}\,E \left( k \right) \propto \left( \left| \gamma \right| - \left| \Delta \right| \right)^{1/2}$ for $\left| \gamma \right| \geq \left| \Delta \right|$.

\begin{figure}[t]
\centering
\includegraphics[width=\linewidth]{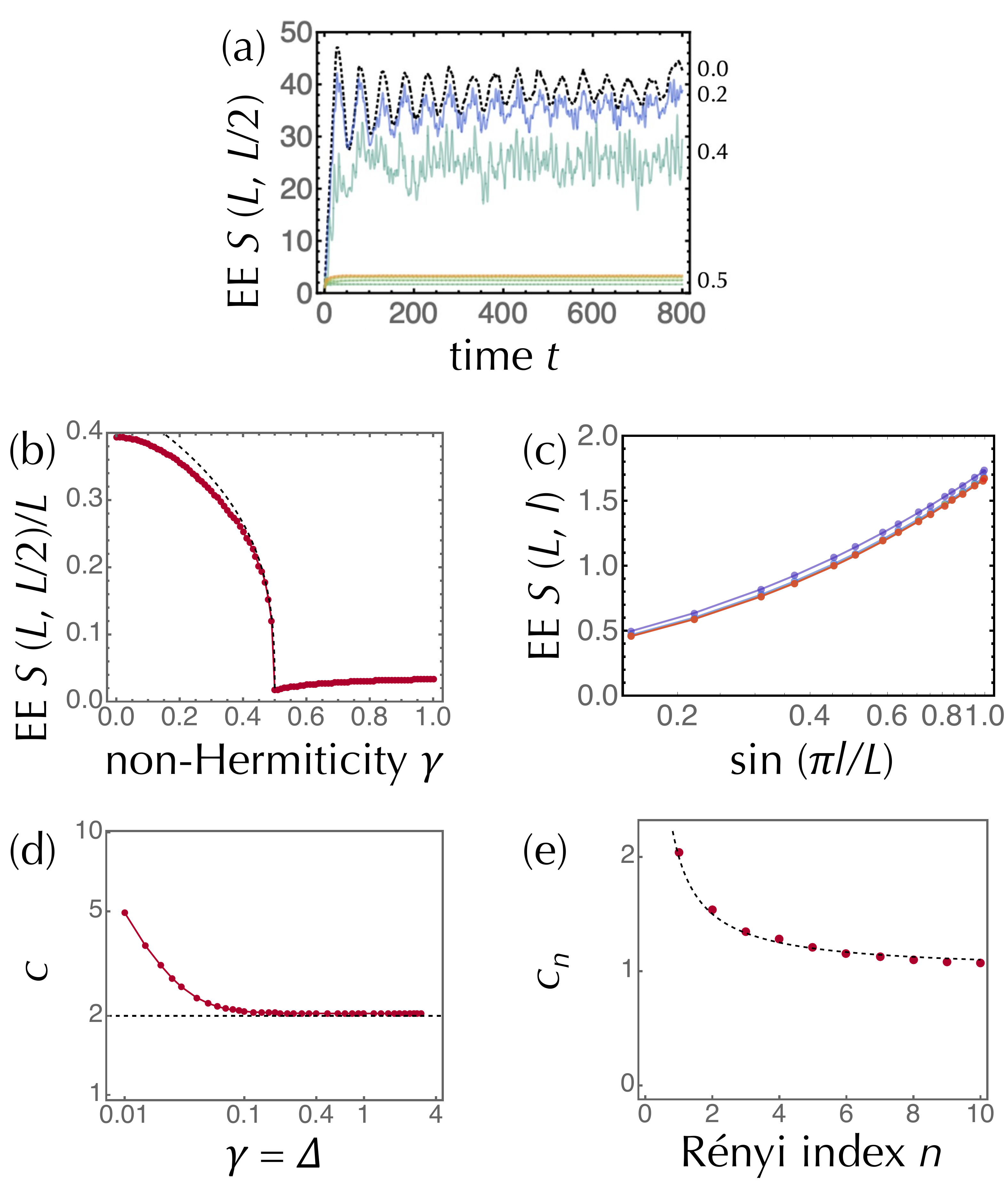} 
\caption{Entanglement entropy of the symplectic Hatano-Nelson model with periodic boundaries ($L=100$, $J=1.0$). 
The initial state is prepared as Eq.~(\ref{eq: sHN-initial-state}).
(a)~Time evolution of the entanglement entropy $S \left( L, L/2 \right)$ ($\Delta = 0.5$) for $\gamma = 0.0$ (black dashed curve), $0.2$ (blue curve), $0.4$, $0.5$, $0.6$ (green curves), $0.8$ (orange curve), and $1.0$ (red curve). 
(b)~Entanglement entropy density $S \left( L, L/2 \right)/L$ for the steady state as a function of non-Hermiticity $\gamma$ ($\Delta=0.5$).
The black dashed curve is the fitting result $S \left( L, L/2 \right)/L = 0.56 \left( \Delta/J - \gamma/J \right)^{0.33}$ around the critical point $\gamma = \Delta$.
(c)~Entanglement entropy $S \left( L, l \right)$ for the steady state at the critical point ($\gamma = \Delta$) as a function of the subsystem length $l$ for $\gamma = 0.0$, $0.1$, $0.2$, $0.4$, $0.6$, $0.8$, and $1.0$. 
(d)~Effective central charge $c$ as a function of $\gamma$ at the critical point ($\gamma = \Delta$). The black dashed line shows $c=2$.
(e)~$c_n$ obtained from the R\'enyi entanglement entropy $S^{(n)} \left( L, l \right)$ for the steady state at the critical point ($\gamma = \Delta = 1.0$) as a function of the R\'enyi index $n$. The black dashed curve shows the conformal field theory result $c_n = c \left( 1+1/n \right)/2$ with $c=2$.
}
	\label{fig: sHN-PBC-EE}
\end{figure}

We also study the entanglement dynamics for the periodic boundary conditions.
Qualitatively, it is similar to the entanglement dynamics for the open boundary conditions:
the entanglement entropy of the nonequilibrium steady state is extensive below the critical point while it is suppressed above the critical point [Fig.~\ref{fig: sHN-PBC-EE}\,(a)].
However, the steady state exhibits a distinct critical behavior around the phase transition point $\left| \gamma \right| = \left| \Delta \right|$.
Below the transition point (i.e., $\left| \gamma  \right| \leq \left| \Delta \right|$), the density of the steady-state entanglement entropy exhibits the critical behavior [Fig.~\ref{fig: sHN-PBC-EE}\,(b)]
\begin{align}
    \frac{S_{\rm s} \left( L, L/2 \right)}{L} \propto \left( \frac{\left| \Delta\right| - \left| \gamma \right|}{J} \right)^{0.33 \pm 0.02},
\end{align}
whose critical exponent $0.33 \pm 0.02$ deviates from that under the open boundary conditions in Eq.~(\ref{eq: sHN-OBC-volume-critical}).
At the critical point $\left| \gamma  \right| = \left| \Delta \right|$, the steady-state entanglement entropy under the periodic boundary conditions is much smaller than that under the open boundary conditions.
According to conformal field theory~\cite{Calabrese-Cardy-04, Calabrese-Cardy-05, *Calabrese-Cardy-06}, the entanglement entropy $S_{\rm s} \left( L, l \right)$ of a one-dimensional quantum critical system with periodic boundaries behaves by
\begin{equation}
    S_{\rm s} \left( L, l \right) = \frac{c}{3} \log \left( \sin \frac{\pi l}{L} \right) + S_0.
        \label{eq: EE-PBC-scaling}
\end{equation}
We confirm that our numerical results for the steady states are consistent with this subextensive behavior [Fig.~\ref{fig: sHN-PBC-EE}\,(c)].
Remarkably, in contrast to the parameter-dependent central charge for the open boundary conditions, the effective central charge does not depend on the system parameter $\left| \gamma/J \right| = \left| \Delta/J \right|$ and is obtained as the following constant [Fig.~\ref{fig: sHN-PBC-EE}\,(d)]:
\begin{equation}
    c = 2.04 \pm 0.08,
        \label{eq: sHN-c-PBC}
\end{equation}
which is compatible with the effective central charge $c=2$ of non-Hermitian free fermions~\cite{Chang-20}.
The different behavior of the effective central charge $c$ means the different universality classes of the entanglement phase transition.
Moreover, we investigate the R\'enyi entanglement entropy for the steady state, which is defined as $S_{\rm s}^{(n)} \coloneqq \left( \mathrm{tr} \log \hat{\rho}^{n} \right)/\left( 1-n \right)$ for the reduced density operator $\hat{\rho}$ and coincides with the von Neumann entanglement entropy $S_{\rm s}$ for $n \to 1$.
According to conformal field theory, the R\'enyi entanglement entropy also follows the scaling in Eq.~(\ref{eq: EE-PBC-scaling}), where the central charge $c$ is replaced by $c_n \coloneqq c \left( 1+1/n \right)/2$~\cite{Calabrese-Cardy-04, Calabrese-Cardy-05, *Calabrese-Cardy-06}.
We also confirm this conformal field theory scaling with respect to the R\'enyi index $n$ [Fig.~\ref{fig: sHN-PBC-EE}\,(e)].
We note that the parameter dependence of the effective central charge for small non-Hermiticity $\gamma$ is due to a finite-size effect that interpolates between the unitary and nonunitary critical points.

Importantly, the mechanism of the entanglement phase transition is different depending on the boundary conditions.
Under the periodic boundary conditions, the entanglement phase transition originates from the real-complex spectral transition.
At the critical point, the Bloch Hamiltonian is defective and exhibits an exceptional point.
This is similar to the entanglement phase transition due to parity-time-symmetry breaking~\cite{Gopalakrishnan-21}.
In such a case, the effective central charge is the constant in Eq.~(\ref{eq: sHN-c-PBC}).
Under the open boundary conditions, on the other hand, the model exhibits no spectral transitions.
While non-Hermiticity is irrelevant to the spectrum, it gives rise to a length scale of the skin modes.
Then, the nonequilibrium quantum criticality is induced by the scale invariance of the skin modes, as discussed in Sec.~\ref{sec: sHN-EP}.
The effective central charge depends on the system parameter [i.e., Eq.~(\ref{eq: sHN-c-OBC})] in contrast to unitary conformal field theory.

Despite these differences, the critical behavior of the bulk modes and that of the boundary (i.e., skin) modes may have a hidden connection with each other.
In fact, the skin effect under the open boundary conditions originates from the non-Hermitian topological invariant under the periodic boundary conditions~\cite{Zhang-20, OKSS-20, KSR-21}, which can be considered as the bulk-boundary correspondence of non-Hermitian topological systems.
In this respect, it is of importance to consider the different critical behaviors of the bulk and boundary modes in terms of nonunitary conformal field theory.
It is also notable that while the bulk and boundary modes are clearly separated in the symplectic Hatano-Nelson model, they can appear simultaneously in more generic non-Hermitian models.

\section{Purification induced by the Liouvillian skin effect}
    \label{sec: Liouvillian}

We have so far considered the conditional dynamics effectively described by non-Hermitian Hamiltonians.
Notably, the skin effect occurs also in the open quantum dynamics described by the master equation~\cite{Breuer-textbook, Rivas-textbook}
\begin{equation}
    \frac{d}{dt} \hat{\rho} = \mathcal{L} \hat{\rho},
\end{equation}
where $\mathcal{L}$ is a Liouvillian that acts on the density operator $\hat{\rho}$ (see Appendix~\ref{asec: NH} for a relationship between non-Hermitian Hamiltonians and Liouvillians in the quantum trajectory approach).
Although the Liouvillian $\mathcal{L}$ is not an operator but a superoperator, it is still non-Hermitian.
Consequently, $\mathcal{L}$ can exhibit the skin effect in a similar manner to non-Hermitian Hamiltonians~\cite{Song-19, Haga-21, Liu-20PRR, Mori-20, Yang-22}.
Here, we demonstrate that the Liouvillian skin effect has a significant influence on the open quantum dynamics described by the master equation.
In particular, we show that the Liouvillian skin effect leads to the purification and the reduction of von Neumann entropy for the steady state.

We consider the following prototypical model that exhibits the Liouvillian skin effect~\cite{Haga-21}:
\begin{align} 
    \mathcal{L} \hat{\rho} 
    \coloneqq 
    \sum_{l=1}^{L} \sum_{n = {\rm R}, {\rm L}} \left( \hat{L}_{ln} \hat{\rho} \hat{L}_{ln}^{\dag} - \frac{1}{2} \{ \hat{L}_{ln}^{\dag} \hat{L}_{ln}, \hat{\rho} \} \right),
        \label{eq: Liouvillian skin}
\end{align}
where the jump operators are
\begin{align}
    \hat{L}_{l{\rm R}} &\coloneqq \sqrt{\frac{J+\gamma}{2}}\,\hat{c}_{l+1}^{\dag} \hat{c}_{l},         
        \label{eq: chiral Liouvillian - R} \\
    \hat{L}_{l{\rm L}} &\coloneqq \sqrt{\frac{J-\gamma}{2}}\,\hat{c}_{l}^{\dag} \hat{c}_{l+1},
        \label{eq: chiral Liouvillian - L}
\end{align}
with $J > 0$ and $\left| \gamma \right| \leq J$.
Similarly to the Hatano-Nelson model, $\hat{L}_{n{\rm R}}$ and $\hat{L}_{n{\rm L}}$ describe the dissipative hopping from the left to the right and from the right to the left, respectively.
Consequently, in the presence of the asymmetry of the hopping (i.e., $\gamma \neq 0$), the spectrum and eigenstates of the Liouvillian dramatically change according to the boundary conditions.
In particular, the steady state $\hat{\rho}_{\rm s}$ greatly depends on the boundary conditions.
In this Liouvillian, the total particle number $\hat{N} = \sum_{l=1}^{L} \hat{c}_{l}^{\dag} \hat{c}_{l}$ is conserved. 
This contrasts with the Liouvillians in Refs.~\cite{Song-19, Liu-20PRR, Yang-22}, in which the total particle number decreases with time.

For the single-particle case, the steady state for the periodic boundary conditions is the completely-mixed state (see Appendix~\ref{asec: Liouvillian} for details)
\begin{equation}
    \hat{\rho}_{\rm s} = \frac{1}{L},
        \label{eq: Liouvillian ss chiral (PBC)}
\end{equation}
while the steady state for the open boundary conditions is the skin modes
\begin{equation}
    \hat{\rho}_{\rm s}
    = \frac{1}{Z} \sum_{l=1}^{L} r^{l} \ket{l} \bra{l}
        \label{eq: Liouvillian ss chiral}
\end{equation}
with $r \coloneqq \left( J+\gamma \right)/\left( J-\gamma \right)$, $\ket{l} \coloneqq \hat{c}_{l}^{\dag} \ket{\rm vac}$, and the normalization constant
\begin{equation}
    Z \coloneqq \sum_{l=1}^{L} r^{L}
    = \frac{r \left( r^{L} - 1 \right)}{r-1}.
\end{equation}
We note in passing that the steady state in Eq.~(\ref{eq: Liouvillian ss chiral}) is formally equivalent to the Gibbs state $Z^{-1} \sum_{l=1}^{L} e^{-\beta E_{l}} \ket{l} \bra{l}$ with the linear potential $\beta E_l \coloneqq -l \log r$.
While the effective temperature is infinite in the absence of the asymmetric hopping (i.e., $\gamma = 0$), it decreases as the asymmetric hopping $\left| \gamma \right|$ increases and reaches zero for the completely-asymmetric hopping $\gamma = \pm J$.

\begin{figure}[t]
\centering
\includegraphics[width=\linewidth]{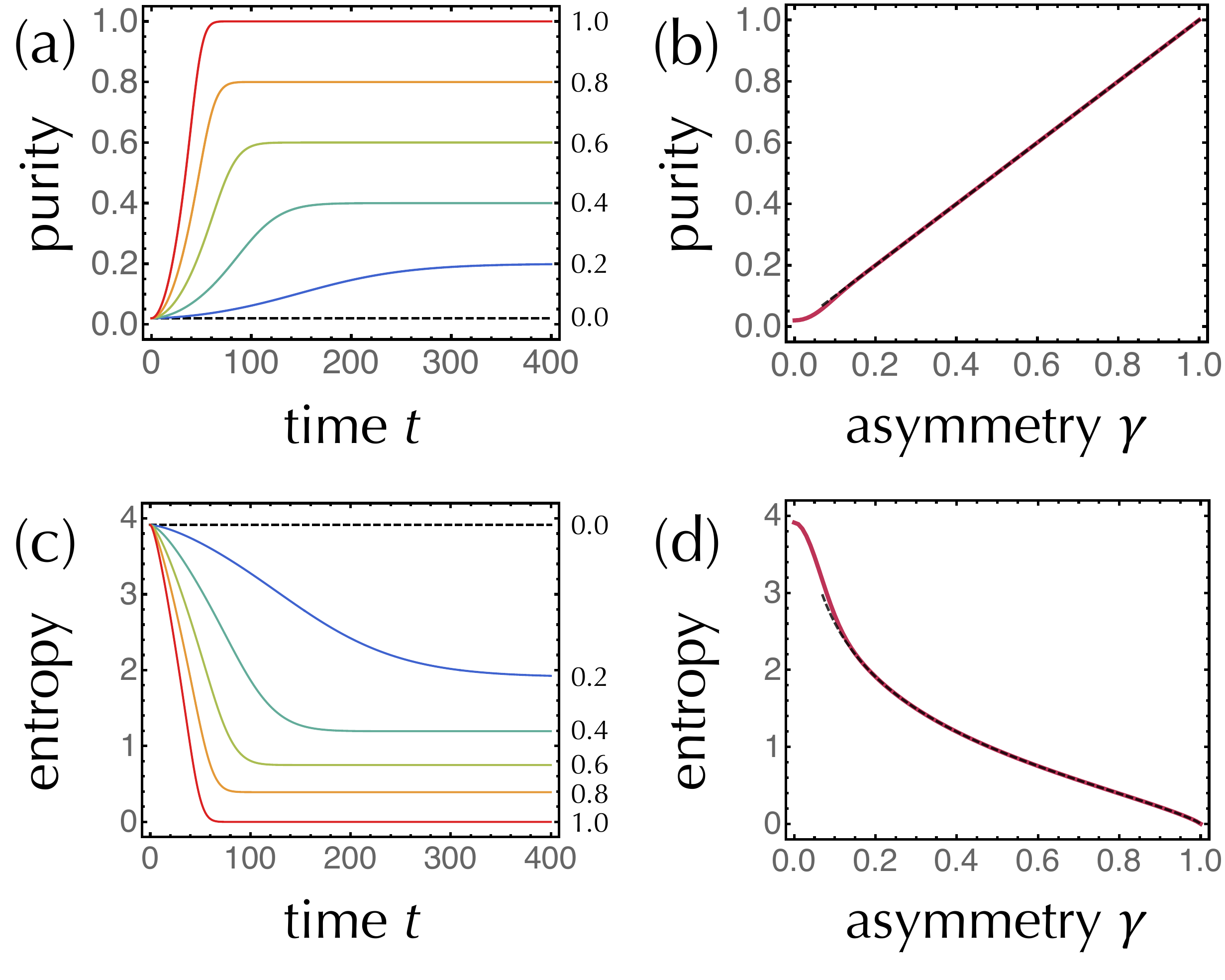} 
\caption{Purification induced by the Liouvillian skin effect ($L=50$, $J=1.0$). The initial state is prepared as the completely-mixed state $\hat{\rho}_0 = 1/L$ with the purity $P_0 = 1/L$ and the von Neumann entropy $S_0 = \log L$. (a)~Time evolution of the purity for $\gamma = 0.0$ (black dashed curve), $\gamma = 0.2$ (blue curve), $\gamma = 0.4$ (green curve), $\gamma = 0.6$ (light-green curve), $\gamma = 0.8$ (orange curve), and $\gamma =1.0$ (red curve). (b)~Steady-state purity as a function of $\gamma$ (red curve), consistent with the analytical result $P_{\rm s} \simeq \gamma/J$ (black dashed curve). (c)~Time evolution of the von Neumann entropy. (d)~Steady-state von Neumann entropy as a function of $\gamma$ (red curve), consistent with the analytical result $S_{\rm s} \simeq \left( J+\gamma/2\gamma \right) \log \left( J+\gamma/2\gamma \right) - \left( J-\gamma/2\gamma \right) \log \left( J-\gamma/2\gamma \right)$ (black dashed curve).}
	\label{fig: Liouvillian-chiral}
\end{figure}

We demonstrate that the skin effect has a considerable influence on the open quantum dynamics even in the Markovian regime.
In particular, the skin effect can purify mixed states.
Let us prepare an initial state as the completely-mixed state $\hat{\rho}_{0} \propto 1$ and consider the dynamics described by the Liouvillian in Eq.~(\ref{eq: Liouvillian skin}).
As shown in Fig.~\ref{fig: Liouvillian-chiral}\,(a), the initially-low purity monotonically increases with time.
The purity for the steady state increases with the larger asymmetry $\left| \gamma \right|$, leading to a pure state for the completely-asymmetric hopping $\gamma = \pm J$ [Fig.~\ref{fig: Liouvillian-chiral}\,(b)].
The steady-state purity is analytically obtained from Eq.~(\ref{eq: Liouvillian ss chiral}) as
\begin{equation}
    P_{\rm s} 
    \coloneqq \mathrm{tr}\,\hat{\rho}_{\rm s}^2
    = \frac{r-1}{r+1} \frac{r^L+1}{r^L-1}
    \simeq \frac{\gamma}{J}
\end{equation}
for $\gamma > 0$ and $L \to \infty$.
This analytical formula is consistent with the numerical results.

We also calculate the time evolution of the von Neumann entropy $S \coloneqq - \mathrm{tr}\,\hat{\rho}_{\rm s} \log \hat{\rho}_{\rm s}$, as shown in Fig.~\ref{fig: Liouvillian-chiral}\,(c).
While the reciprocal dynamics realizes the maximal entropy, the asymmetry of the dissipative hopping lowers the entropy.
The entropy $S_{\rm s}$ for the steady state monotonically decreases as a function of $\left| \gamma \right|$, reaching zero for the completely-asymmetric case $\gamma = \pm J$ [Fig.~\ref{fig: Liouvillian-chiral}\,(d)].
Here, $S_{\rm s}$ is also analytically obtained as
\begin{align}
    S_{\rm s} 
    &\coloneqq - \mathrm{tr}\,\hat{\rho}_{\rm s} \log \hat{\rho}_{\rm s} \nonumber \\
    &= \log Z - \frac{\log r}{r-1} \frac{Lr^{L+1}}{Z} + \frac{\log r}{r-1} \nonumber \\
    &\simeq \left( \frac{J+\gamma}{2\gamma} \right) \log \left( \frac{J+\gamma}{2\gamma} \right) - \left( \frac{J-\gamma}{2\gamma} \right) \log \left( \frac{J-\gamma}{2\gamma} \right)
\end{align}
for $\gamma > 0$ and $L \to \infty$.
Notably, while the steady-state entropy $S_{\rm s}$ subextensively increases with respect to the system length $L$ (i.e., $S_{\rm s} = \log L$) in the absence of the skin effect, $S_{\rm s}$ is independent of $L$ (i.e., area law) in the presence of the skin effect.
This is similar to the entanglement suppression of the open quantum dynamics effectively described by a non-Hermitian Hamiltonian that is discussed in the previous sections.

The purification and suppression of the von Neumann entropy are induced by the Liouvillian skin effect.
Under the periodic boundary conditions, no skin effect occurs, and the steady state is the completely-mixed state in Eq.~(\ref{eq: Liouvillian ss chiral (PBC)}).
Consequently, no purification occurs, and the steady state is characterized by the maximal entropy.

It is also notable that purification can arise from quantum measurements~\cite{Chan-19, Skinner-19, Li-18, *Li-19, Choi-20, *Bao-20, Gullans-20, Cao-19, Jian-20, Lavasani-21, Sang-21, Ippoliti-21, Fuji-20, Alberton-21, Ippoliti-21-spacetimeL, *Ippoliti-21-spacetimeX, Lu-21}.
However, such measurement-induced purification occurs only in the conditional dynamics of a particular quantum trajectory.
This conditional nature of the open quantum dynamics is a key to the measurement-induced phase transitions.
By contrast, we here demonstrate that the skin effect leads to the purification even in the Markovian master equation characterized by a Liouvillian, which describes the open quantum dynamics averaged over multiple quantum trajectories.
This also shows a significant role of the skin effect in the open quantum dynamics.

\section{Discussions}
    \label{sec: conclusion}

The entanglement dynamics provides the foundations of quantum statistical physics.
However, the nature of entanglement in open quantum systems has remained elusive in contrast to closed quantum systems.
In this work, we have shown that the skin effect, 
a universal feature intrinsic to non-Hermitian systems, has a significant impact on the entanglement dynamics in open quantum systems.
We have shown that the skin effect suppresses the entanglement growth and even induces an entanglement phase transition.
This is different from the known mechanism that triggers the entanglement phase transition such as quantum measurements~\cite{Chan-19, Skinner-19, Li-18, *Li-19, Choi-20, *Bao-20, Gullans-20, Cao-19, Jian-20, Lavasani-21, Sang-21, Ippoliti-21, Fuji-20, Alberton-21, Ippoliti-21-spacetimeL, *Ippoliti-21-spacetimeX, Lu-21}.
While we have considered the prototypical models for illustrative purposes, the skin effect originates solely from non-Hermitian topology, and hence our entanglement phase transition should generally arise in a wide range of open quantum systems.
On the basis of the recent experimental observations of the skin effect in open quantum systems~\cite{Xiao-19-skin-exp, Liang-22}, as well as the realization of non-Hermitian spin-orbit-coupled fermions~\cite{Ren-22}, our results should be observed in a similar experimental setup.

We have shown that our entanglement phase transition accompanies anomalous nonequilibrium quantum criticality that is described by the boundary-sensitive effective central charges 
[cf. the difference between Figs.~\ref{fig: sHN-EE-critical}\,(b) and \ref{fig: sHN-PBC-EE}\,(d)].
These anomalous critical behaviors imply a new universality class of phase transitions in open quantum systems.
It merits further study to derive the nonunitary conformal field theory that describes the nonequilibrium quantum criticality induced by the skin effect.
The different critical behaviors in the bulk and boundaries may be unified into the same field theory.
In this respect, it is worth noting that the skin effect can be considered as a quantum anomaly of a topological field theory intrinsic to non-Hermitian systems~\cite{KSR-21}.

Furthermore, we have demonstrated that our entanglement phase transition is induced by the criticality of skin modes that decay according to the power law.
Notably, while the conventional Bloch band theory cannot describe the skin modes, recent works developed a non-Bloch band theory that correctly characterizes the skin modes~\cite{YW-18-SSH, Yokomizo-19, KOS-20}.
However, the non-Bloch band theory only predicts the exponentially-localized skin modes and cannot describe the critical skin modes discovered in this work.
It is significant to generally develop a modified band theory that captures the phase transitions and critical phenomena induced by the non-Hermitian skin effect.
Additionally, the skin effect leads to the slowdown of relaxation processes~\cite{Haga-21}.
The critical skin effect should yield the logarithmic correction of the relaxation time.

We have also shown that the skin effect plays an important role in the open quantum dynamics described by the master equation.
In particular, the skin effect changes the properties of the nonequilibrium steady state and increases the purity and decreases the von Neumann entropy.
These findings may lead to potential applications of the skin effect in quantum information science.
They also imply that the skin effect has a considerable impact in a wide range of open classical and quantum dynamics.
In this research direction, it is worth studying the role of the skin effect, for example, in quantum circuits.
We note in passing that recent works have found signatures of non-Hermitian topology in monitored quantum circuits~\cite{Kells-21, Fleckenstein-22}.
Moreover, it is meaningful to explore the relevance of the skin effect in classical stochastic processes such as the asymmetric simple exclusion process~\cite{Derrida-review}.

Another remarkable mechanism that prohibits the quantum diffusion is disorder.
In closed quantum systems, sufficiently strong disorder drives the systems into the Anderson~\cite{Lee-review, Evers-review} or many-body~\cite{Huse-review} localization, resulting in the absence of thermalization.
While the skin effect also accompanies an extensive number of localized eigenmodes similarly to the disorder-induced localization, we emphasize that it does not rely on disorder and thus gives a different mechanism that hinders the entanglement propagation and thermalization.
Meanwhile, it is intriguing to consider the open quantum dynamics in the presence of disorder.
In fact, non-Hermiticity changes the universality classes of localization transitions~\cite{Hatano-Nelson-96, *Hatano-Nelson-97, Efetov-97, Feinberg-97, Brouwer-97, Hamazaki-19, Longhi-19, Tzortzakakis-20, Huang-20, KR-21, Luo-21L, *Luo-21B, *Luo-22R}.
The interplay of disorder and dissipation should further enrich phase transitions and critical phenomena in open quantum systems.

While we have focused on one-dimensional systems in this work, it is also worthwhile to study non-Hermitian systems in higher dimensions.
Different types of skin effects appear in higher dimensions, such as the chiral magnetic skin effect~\cite{Terrier-20, Bessho-20, Denner-21, KSR-21}, higher-order skin effect~\cite{Okugawa-20, KSS-20}, 
and defect-induced skin effect~\cite{Sun-21, Schindler-21, Bhargava-21}.
These higher-dimensional skin effects may give rise to further different universality classes of phase transitions and critical phenomena in open quantum systems.
It is also of interest to study the entanglement dynamics of non-Hermitian interacting systems.
Several recent works have shown that the interplay of non-Hermiticity and many-body interactions leads to new quantum phases~\cite{Guo-20, Yoshida-19, Mu-20, Xi-19, Lee-20, Matsumoto-21, Liu-20, Yang-Morampudi-QSL-21}.
Similarly to the many-body localized phases due to disorder~\cite{Znidaric-08, Bardarson-12, Serbyn-13}, many-body skin modes may exhibit the logarithmic violation of the area law for the entanglement growth.

\section*{Acknowledgements}
We thank Anish Kulkarni and Yuhan Liu for helpful discussions.
K.K. is supported by the Japan Society for the Promotion of Science (JSPS) through the Overseas Research Fellowship.
S.R. is supported by the National Science Foundation under award number DMR-2001181, and by a Simons Investigator Grant from the Simons Foundation (Award Number: 566116).
This work is supported by the Gordon and Betty Moore Foundation through Grant GBMF8685 toward the Princeton theory program.

\appendix

\section{Effective non-Hermitian Hamiltonians}
    \label{asec: NH}

The non-Hermitian Hamiltonians in Eqs.~(\ref{eq: HN}) and (\ref{eq: sHN}) can be realized in the quantum trajectory approach~\cite{Dalibard-92, *Molmer-93, Dum-92, Carmichael-textbook, Plenio-review, Daley-review}.
Let us consider a Markovian open quantum system, which is generally described by the Lindblad master equation~\cite{Breuer-textbook, Rivas-textbook}:
\begin{equation}
    \frac{d}{dt} \hat{\rho}
    = -\ii\,[ \hat{H}, \hat{\rho} ]
    + \sum_{n} \left( \hat{L}_{n} \hat{\rho} \hat{L}_{n}^{\dag} - \frac{1}{2} \{ \hat{L}_{n}^{\dag} \hat{L}_{n}, \hat{\rho} \}\right),
\end{equation}
where $\hat{\rho}$ is the density operator, $\hat{H}$ is the Hamiltonian that describes the coherent dynamics, and $\hat{L}_n$'s are the jump operators that describe the coupling to the external environment.
This master equation can be written as
\begin{equation}
    \frac{d}{dt} \hat{\rho}
    = -\ii \left( \hat{H}_{\rm eff} \hat{\rho} - \hat{\rho} \hat{H}_{\rm eff}^{\dag} \right)
    + \sum_{n} \hat{L}_{n} \hat{\rho} \hat{L}_{n}^{\dag} 
\end{equation}
with the effective non-Hermitian Hamiltonian
\begin{equation}
    \hat{H}_{\rm eff} \coloneqq \hat{H} - \frac{\ii}{2} \sum_{n} \hat{L}_{n}^{\dag} \hat{L}_{n}.
\end{equation}
The last term $\sum_{n} \hat{L}_{n} \hat{\rho} \hat{L}_{n}^{\dag}$ specifies each quantum trajectory subject to stochastic loss events.
Here, $\hat{L}_{n} \sqrt{dt}$ can be considered to be a measurement operator for a signal $n$ in the time interval $\left[ t, t+dt \right]$, and $1-\ii \hat{H}_{\rm eff} dt$ can be considered to be a measurement operator for no signals.
Under continuous monitoring and postselection of the null measurement outcome, the quantum jumps are no longer relevant, and the dissipative dynamics is described by the effective non-Hermitian Hamiltonian $\hat{H}_{\rm eff}$.

To obtain the Hatano-Nelson model in Eq.~(\ref{eq: HN}), we choose the Hamiltonian $\hat{H}$ and the jump operators $\hat{L}_{l}$'s ($l = 1, 2, \cdots, L$) to be~\cite{Gong-18}
\begin{align}
    \hat{H} &= - \frac{J}{2} \sum_{l=1}^{L} \left( \hat{c}_{l+1}^{\dag} \hat{c}_{l} 
    + \hat{c}_{l}^{\dag} \hat{c}_{l+1} \right), \\
    \hat{L}_{l} &= \sqrt{\left| \gamma \right|} \left( \hat{c}_{l} + \ii\,\mathrm{sgn} \left( \gamma \right) \hat{c}_{l+1} \right).
\end{align}
Although the effective Hamiltonian $\hat{H}_{\rm eff}$ differs from Eq.~(\ref{eq: HN}) by the background constant loss $-\ii \left| \gamma \right| \hat{N} = -\ii \left| \gamma \right| \sum_{l=1}^{L} \hat{c}_{l}^{\dag} \hat{c}_{l}$, it only describes the total decay of the system and does not contribute to the dynamics of the wave function.
Similarly, to obtain the symplectic Hatano-Nelson model in Eq.~(\ref{eq: sHN}), we choose $\hat{H}$ and $\hat{L}_{l}$'s ($l = 1, 2, \cdots, L$) to be
\begin{align}
    \hat{H} &= - \frac{1}{2} \sum_{l=1}^{L} \left[ \hat{c}_{l+1}^{\dag} \left( J-\ii\Delta\sigma_x \right)  \hat{c}_{l} \right. \nonumber \\ 
    &\qquad\qquad\qquad \left. + \hat{c}_{l}^{\dag} \left( J+\ii\Delta\sigma_x \right) \hat{c}_{l+1} \right], \\
    \hat{L}_{l, \uparrow} &= \sqrt{\left| \gamma \right|} \left( \hat{c}_{l} + \ii\,\mathrm{sgn} \left( \gamma \right) \hat{c}_{l+1} \right), \\
    \hat{L}_{l, \downarrow} &= \sqrt{\left| \gamma \right|} \left( \hat{c}_{l} - \ii\,\mathrm{sgn} \left( \gamma \right) \hat{c}_{l+1} \right).
\end{align}

As described above, the open quantum dynamics characterized by the non-Hermitian Hamiltonian is conditional, and the success probability of having the desirable non-Hermitian Hamiltonian can be low at long time.
This is different from the quantum master equation, which describes the open quantum dynamics of the mixed states averaged over many quantum trajectories and hence is free from the postselection.
However, in certain cases, this difficulty can be circumvented, and the effective non-Hermitian Hamiltonian is well realized with a reasonable probability (see, for example, Ref.~\cite{Lee-14X, Lee-14L}).
In this respect, it is also notable that a similar experimental difficulty should arise also in the measurement-induced phase transitions.
In fact, only a quantum trajectory conditioned on a set of measurement outcomes can exhibit an entanglement phase transition, while the mixed quantum state averaged over many quantum trajectories should not exhibit such a phase transition.
Still, a different way to realize the measurement-induced phase transitions without the postselection of a certain set of measurement outcomes has recently been proposed~\cite{Ippoliti-21-spacetimeL}.
Finally, while we here focus on the quantum trajectory approach, it should be noted that the effective non-Hermitian Hamiltonians can be justified also by the Feshbach projection formalism~\cite{Gamow-28, Feshbach-54, *Feshbach-58, *Feshbach-62, Rotter-review, Moiseyev-textbook}.

\section{Numerical method for the dynamics of non-Hermitian free fermions}
    \label{asec: numerics}

We describe a numerical method to investigate the dynamics of non-Hermitian free (i.e., quadratic) fermions.
An initial state $\ket{\psi_0}$ evolves by the non-Hermitian Hamiltonian $\hat{H}$ as Eq.~(\ref{eq: NH-dynamics}).
The denominator $\| e^{-\ii \hat{H} t} \ket{\psi_0} \|$ describes the normalization of the evolved state due to the conditional nature of the non-Hermitian Hamiltonian.
This time evolution is equivalently described by the nonlinear Schr\"odinger equation~\cite{Brody-12}
\begin{equation}
    \ii \frac{d}{dt} \ket{\psi} 
    = \left( \hat{H} - \braket{\psi | \hat{H} | \psi} \right) \ket{\psi}.
\end{equation}
Despite non-Hermiticity of the Hamiltonian, the total particle number is conserved under the dynamics when the initial state is an eigenstate of the particle number operator (i.e., $\hat{N} \ket{\psi_0} = N \ket{\psi_0}$).
This is a consequence of $\mathrm{U} \left( 1 \right)$ symmetry $[ \hat{H}, \hat{N} ] = 0$.

We first consider a spinless-fermionic system such as the Hatano-Nelson model.
We prepare an initial state as a Gaussian state with a fixed particle number $N$.
As an advantage of the quadratic Hamiltonian, the evolved state remains to be a Gaussian state through the time evolution in Eq.~(\ref{eq: NH-dynamics}).
Thus, the state can always be represented as
\begin{equation}
    \ket{\psi \left( t \right)}
    = \prod_{n=1}^{N} \left( \sum_{l=1}^{L} U_{ln} \left( t \right) \hat{c}_{l}^{\dag} \right) \ket{\vac},
\end{equation}
where $\ket{\mathrm{\vac}}$ is the fermionic vacuum and $U$ is the $L \times N$ isometry satisfying
\begin{equation}
    U^{\dag} U = 1.
\end{equation}
In this representation, the matrix $U = U \left( t \right)$ contains all information about the quantum dynamics.
In particular, the $L \times L$ correlation matrix
\begin{equation}
    C_{ij} \left( t \right)
    \coloneqq \braket{\psi \left( t \right) | \hat{c}_{i}^{\dag} \hat{c}_{j} | \psi \left( t \right)}
\end{equation}
is obtained as
\begin{equation}
    C \left( t \right)
    = \left[ U\left( t \right) U^{\dag} \left( t \right) \right]^{T}.
\end{equation}
From the correlation matrix, the local particle number in Eq.~(\ref{eq: HN - local particle number}) reads
\begin{equation}
    n_{l} \left( t \right) = C_{ll} \left( t \right),
\end{equation}
and the local charge current in Eq.~(\ref{eq: HN - local current}) reads
\begin{equation}
    I_{l} \left( t \right) = J\,\mathrm{Im} \left[ C_{l+1, l} \left( t \right) \right].
\end{equation}
To calculate the entanglement entropy $S$ between the subsystem $[ x_1, x_2 ]$ and the rest of the system, we diagonalize the $\left( x_2 - x_1 + 1 \right) \times \left( x_2 - x_1 + 1 \right)$ submatrix $\left[ C \right]_{i, j = x_1}^{x_2}$ and obtain its eigenvalues $\lambda_n$'s ($n = 1, 2, \cdots, x_2 - x_1 + 1$).
Then, the von Neumann entanglement entropy is given as
\begin{equation}
    S = - \sum_{i=1}^{x_2 - x_1 + 1} \left[ \lambda_i \log \lambda_i + \left( 1 - \lambda_i \right) \log \left( 1 - \lambda_i \right) \right],
\end{equation}
and the R\'enyi entanglement entropy is  
\begin{equation}
    S^{(n)} = \frac{1}{1-n} \sum_{i=1}^{x_2 - x_1 + 1} \log \left[ \lambda_i^n + \left( 1 - \lambda_i \right)^n \right].
\end{equation}
with the R\'enyi index $n$.
Here, we calculate the entanglement entropy from a single wave function instead of the biorthogonal density operator constructed from both right and left eigenstates~\cite{Couvreur-17, Herviou-19-ES, Chang-20}.

The time evolution of $U = U \left( t \right)$ is efficiently calculated as follows.
After the time interval $\Delta t$, the state evolves as
\begin{align}
    &\ket{\psi \left( t+\Delta t \right)}
    \propto e^{-\ii \hat{H}\Delta t} \ket{\psi \left( t \right)} \nonumber \\
    &\qquad = \prod_{n=1}^{N} \left( \sum_{l=1}^{L} \left[ e^{-\ii h\Delta t} U \right]_{ln} \left( t \right) \hat{c}_{l}^{\dag} \right) \ket{\vac},
\end{align}
where $h$ is the $L \times L$ single-particle Hamiltonian (i.e., $\hat{H} = \sum_{i, j = 1}^{L} \hat{c}_{i}^{\dag}  h_{ij} \hat{c}_{j}$).
To restore the normalization condition $\braket{\psi \left( t \right) | \psi \left( t \right)} = 1$, we perform the QR decomposition
\begin{equation}
    e^{-\ii h\Delta t} U = QR,
\end{equation}
where $Q$ is an $L \times N$ matrix satisfying $Q^{\dag} Q = 1$ and $R$ is an upper triangular matrix.
The $L \times N$ matrix $U \left( t+\Delta t \right)$ is obtained as 
\begin{equation}
    U \left( t+\Delta t \right) = Q.
\end{equation}
In our numerical calculations, we choose $\Delta t = 0.05$ for $J=1.0$.
This numerical method is applicable even in the presence of spatial or temporal disorder.
A similar numerical method was used to investigate the open quantum dynamics of monitored free fermions~\cite{Cao-19, Alberton-21}.

The dynamics of a spinful system including the symplectic Hatano-Nelson model in Eq.~(\ref{eq: sHN}) can also be calculated in a similar manner.
In the spinful case, the state is represented as
\begin{equation}
    \ket{\psi \left( t \right)}
    = \prod_{n=1}^{N} \left( \sum_{l=1}^{L} \sum_{s=\uparrow, \downarrow} U_{lsn} \left( t \right) \hat{c}_{ls}^{\dag} \right) \ket{\vac},
\end{equation}
where $s$ describes the spin degree of freedom, and the isometry $U$ is now the $2L \times N$ matrix satisfying $U^{\dag} U = 1$.
From $U$, the $2L \times 2L$ correlation matrix $C$ is obtained as
\begin{align}
    C_{is, js'} \left( t \right)
    &\coloneqq \braket{\psi \left( t \right) | \hat{c}_{is}^{\dag} \hat{c}_{js'} | \psi \left( t \right)} \nonumber \\
    &= \left[ U\left( t \right) U^{\dag} \left( t \right) \right]_{js', is}.
\end{align}

\section{Different initial conditions}
    \label{asec: initial conditions}

\begin{figure}[t]
\centering
\includegraphics[width=\linewidth]{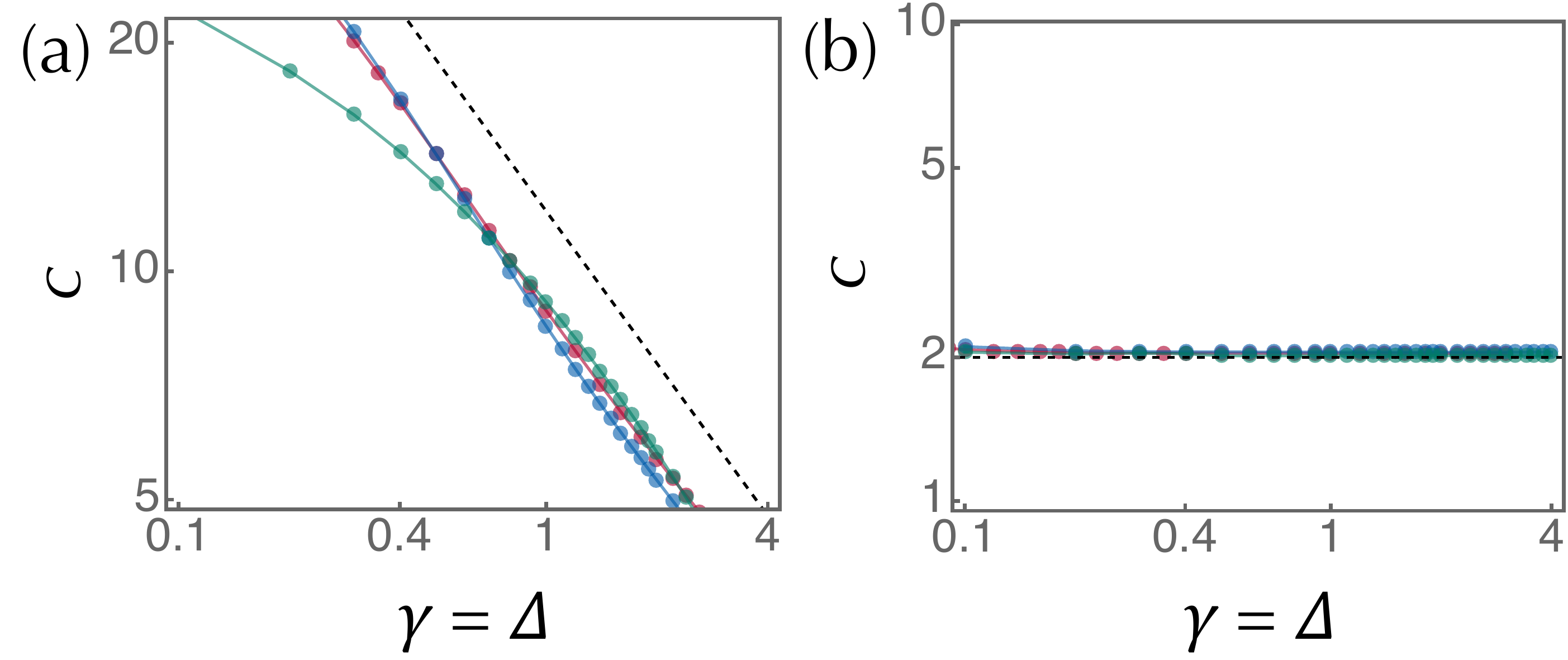} 
\caption{Effective central charge $c$ of the symplectic Hatano-Nelson model ($L=100$, $J=1.0$) at the critical point ($\gamma = \Delta$) under the (a)~open boundary conditions and (b)~periodic boundary conditions. 
For each $\gamma = \Delta$, the effective central charge $c$ is obtained from the logarithmic scaling of the steady-state entanglement entropy for the initial states in Eq.~(\ref{eq: sHN-initial-state}) (red dots), Eq.~(\ref{aeq: sHN-initial-state (2)}) (blue dots), and Eq.~(\ref{aeq: sHN-initial-state (3)}) (green dots).
The black dashed lines are (a)~$c \propto \gamma^{-2/3}$ and (b)~$c=2$.}
	\label{afig: sHN-EE-critical}
\end{figure}

We provide additional numerical results on the critical behavior in the symplectic Hatano-Nelson model for different initial conditions.
We prepare the initial state as the fully polarized state
\begin{equation}
    \ket{\psi_0} = \left( \prod_{l=1}^{L} \hat{c}_{l, \uparrow}^{\dag}\right) \ket{\rm vac}
        \label{aeq: sHN-initial-state (2)}
\end{equation}
and obtain the effective central charge from the logarithmic scaling of the steady-state entanglement entropy for both open and periodic boundary conditions (Fig.~\ref{afig: sHN-EE-critical}).
The obtained effective central charges are consistent with those for the different initial state in Eq.~(\ref{eq: sHN-initial-state}).
We also prepare the initial state as
\begin{equation}
    \ket{\psi_0} = \left( \prod_{l=1}^{L/4} \hat{c}_{4l-3, \uparrow}^{\dag} \hat{c}_{4l-3, \downarrow}^{\dag} \right) \ket{\rm vac},
        \label{aeq: sHN-initial-state (3)}
\end{equation}
which has the different particle number.
Under the open boundary conditions, the effective central charges behave differently for $\left| \gamma \right| \ll J$, in which the universal behavior should not be expected because of a significant crossover between the unitary and nonunitary critical points.
For $\left| \gamma \right| \simeq J$, on the other hand, the power-law scaling $c \propto \gamma^{-2/3}$ in Eq.~(\ref{eq: sHN-c-OBC}) appears.
Under the periodic boundary conditions, the effective central charge is obtained as $c \simeq 2$ and hence consistent with those for the different initial conditions.

\section{Diagonalization of Liouvillians}
    \label{asec: Liouvillian}

We exactly solve the Liouvillian described by Eqs.~(\ref{eq: Liouvillian skin}), (\ref{eq: chiral Liouvillian - R}), and (\ref{eq: chiral Liouvillian - L}) in the single-particle Hilbert space~\cite{Haga-21}.
First, for the periodic boundary conditions, we have
\begin{align}
    \mathcal{L} \ket{l} \bra{l}
    &= \frac{1}{2} \left[ \left( J+\gamma \right) \ket{l+1} \bra{l+1} \right. \nonumber \\
    &\qquad\quad\left. + \left( J-\gamma \right) \ket{l-1} \bra{l-1} \right] 
    - J \ket{l} \bra{l}
        \label{aeq: Liouvillian chiral eigenequation}
\end{align}
for $l=1, 2, \cdots, L$.
Here, $\ket{l} \coloneqq \hat{c}_{l}^{\dag} \ket{\rm vac}$ is the single-particle state at site $l$, and we have $\ket{0} = \ket{L}$ and $\ket{L+1} = \ket{1}$ owing to the periodic boundary conditions.
Notably, Eq.~(\ref{aeq: Liouvillian chiral eigenequation}) is formulated in the subspace spanned solely by the diagonal states $\left\{ \ket{1} \bra{1}, \ket{2} \bra{2}, \cdots, \ket{L} \bra{L} \right\}$.
The matrix representation of $\mathcal{L}$ in this subspace coincides with the single-particle matrix of the Hatano-Nelson model in Eq.~(\ref{eq: HN}) with periodic boundaries.
Therefore, the eigenvalues of $\mathcal{L}$ are 
\begin{align}
    &\frac{1}{2} \left[ \left( J+\gamma \right) e^{-\ii k} + \left( J-\gamma \right) e^{\ii k} \right] - J \nonumber \\
    &\qquad\qquad\qquad\qquad= J \left( \cos k - 1 \right) - \ii \gamma \sin k,
\end{align}
and the corresponding eigenstates are the plane waves
\begin{align}
    \frac{1}{L} \sum_{l=1}^{L} e^{\ii kl} \ket{l} \bra{l}
\end{align}
with $k \in \{ 0, 2\pi/L, 4\pi/L, \cdots, 2\pi \left(L-1 \right)/L\}$.
Thus, the steady state, which is the zero mode of $\mathcal{L}$, is given as the plane wave with zero momentum $k=0$:
\begin{equation}
    \hat{\rho}_{\rm s} = \frac{1}{L} \sum_{l=1}^{L} \ket{l} \bra{l}.
\end{equation}
The other eigenstates superposed by off-diagonal states do not contribute to the steady state~\cite{Haga-21}.

For the open boundary conditions, on the other hand, the Liouvillian exhibits the skin effect in a similar manner to the Hatano-Nelson model.
We still have Eq.~(\ref{aeq: Liouvillian chiral eigenequation}) for the bulk $l=2, 3, \cdots, L-1$.
At the boundaries, we have
\begin{align}
    \mathcal{L} \ket{1} \bra{1}
    &= \frac{J+\gamma}{2} \ket{2} \bra{2} - \frac{J+\gamma}{2} \ket{1} \bra{1}, \\
    \mathcal{L} \ket{L} \bra{L}
    &= \frac{J-\gamma}{2} \ket{L-1} \bra{L-1} - \frac{J-\gamma}{2} \ket{L} \bra{L}.
\end{align}
Because of the different boundary conditions, the steady state of $\mathcal{L}$ is now given as
\begin{equation}
    \hat{\rho}_{\rm s} \propto
    \sum_{l=1}^{L} r^{l} \ket{l} \bra{l}
\end{equation}
with $r \coloneqq \left( J+\gamma \right)/\left( J-\gamma \right)$.

It is also notable that the relaxation process speeds up for the larger asymmetry $\gamma$.
The relaxation time $\tau$ subject to the Liouvillian skin effect is given as
\begin{equation}
    \tau \simeq \frac{L}{\xi \Delta},
\end{equation}
where $\xi$ is the localization length of the skin mode, and $\Delta$ is the Liouvillian gap~\cite{Haga-21}.
From $\xi = 1/\log r$ and $\Delta = J - \sqrt{J^2 - \gamma^2}$ for Eq.~(\ref{eq: Liouvillian skin}), we have
\begin{equation}
    \tau \simeq \frac{L}{J - \sqrt{J^2-\gamma^2}} \log \frac{J+\gamma}{J-\gamma}.
\end{equation}
This is a decreasing function of $0 \leq \gamma \leq J$ and consistent with the numerical results in Fig.~\ref{fig: Liouvillian-chiral}.

\bibliography{skin-entanglement-dynamics}

\end{document}